\documentclass[10pt,fleqn]{article}
\makeatletter
  \def\eqalign#1{\null\,\vcenter{\openup\jot\m@th
    \ialign{\strut\hfil$\displaystyle{##}$&$\displaystyle{{}##}$\hfil
      \crcr#1\crcr}}\,}
\def\cmeqalign#1{\null\,\vcenter{\openup\jot\m@th
  \ialign{\strut\hfil$\displaystyle{##}$&&\hfil$\displaystyle{{}##}$\hfil
      \crcr#1\crcr}}\,}
\makeatother
\let\Message=\typeout
\def\half{{\displaystyle {1\over2}}}
\def\grad{\mathop{\rm grad}\nolimits}

\def\oversymbol#1#2{\vbox{\ialign{##\crcr \hfil$#1$\hfil\crcr
   \noalign{\kern1pt\nointerlineskip}%
   \hbox{$\hfil\displaystyle#2\hfil$}\crcr}}}
\def\overeq#1{\oversymbol{\scriptstyle\kern.5pt =}{#1}}
\def\dualp#1{{}^{\ast_{(\hbox{$\scriptstyle #1$})}} \kern-1pt}

\def\bivec#1{\vbox{\ialign{##\crcr $\leftrightarrow$\crcr\noalign{
   \kern-1pt \nointerlineskip}$\hfil\displaystyle{#1}\hfil$\crcr}}}
  
   \def\pub(#1){(#1)}
\def\be{\begin{equation}}
\def\ee{\end{equation}}

\def\Sum{\sum\limits}
\def\defin{\stackrel{\rm def}{=}\ }
\def\vainf#1{\raisebox{-1.5ex}{$\stackrel {\displaystyle {\longrightarrow } }
        {\scriptscriptstyle{#1\rightarrow\infty }} $}}

\def\llongrightarrow{{-\!\!\!-\!\!\!-\!\!\!\!\longrightarrow}}
\def\maggord{\raisebox{-.8ex}{$\ \stackrel{>}{\sim}\ $} }
\def\minord{\raisebox{-.8ex}{$\ \stackrel{<}{\sim}\ $} }

\def\p{{{\bf p}}}
\def\q{{{\bf q}}}
\def\u{{{\bf u}}}

\def\x{{{\bf x}}}
\def\y{{{\bf y}}}

\def\D{{{\bf D}}}

\def\bfxi{{\mbox{\boldmath$\xi$}}}
\def\bfnu{{\mbox{\boldmath$\nu$}}}

\def\Ucal{{\cal U}}

\def\Wcal{{\cal W}}
\def\Ncal{{\cal N}}
\def\Hcal{{\cal H}}
\def\Rcal{{\cal R}}
\def\RcA{{\Rcal^{\!\!{}^{\scriptscriptstyle[A]}}}\!}
\def\Rc0{{\Rcal^{\!\!{}^{\scriptscriptstyle[0]}}}\!}
\def\bfHcal{{\mbox{\boldmath$\Hcal$}}}
\def\bfRcal{{\mbox{\boldmath$\Rcal$}}}
\def\bfRcA{{\mbox{\boldmath$\RcA$}}}
\def\Gabc{\Gamma^a_{bc}}

\def\ric{{\sf Ric}}
\def\s|{\left|\!\left|}
\def\d|{\right|\!\right|}

\def\qquad{\  \  \  \  }
\def\jlc{{{\sf JLC~}}}
\def\fpub{{{\sf FPU-}$\!\beta\! $~ }}

\def\eg{{\tt e.g.}}
\def\ie{{\rm i.e.}}
\def\lcn{{\sf LCN}}
\def\LCN{{\sf LCN~}}
\def\mdf{{\sf mdf~}}
\def\SM{{\sf SM~}}
\def\DS{{\sf DS}}
\def\GDA{{\sf GDA}}
\def\GDI{{{\sf GDI}'s}}
\def\FPU{{\sf FPU~}}
\def\rhs{{r.h.s.~}}

\def\SST{{\sf SST}}
\def\TqM{{{\sf T}_\q{\sf M}}}
\begin{document}
\begin{titlepage}
\baselineskip20pt
\setcounter{page}{1}
\begin{center} 
{\LARGE 
{\bf {Dynamical Instability\\ and Statistical behaviour\\ 
of N-bodySystems.\\}}
}
\vskip3.truecm 
{\Large 
Piero CIPRIANI$^{1,2\dagger}$\\
Maria DI BARI$^{1\ast}$\\
} 
\vskip.8truecm 
{\large
$^1${Dipartimento di Fisica {\it "E. Amaldi"}, Universit\`a 
{\it "Roma Tre"},\\
Via della Vasca Navale, 84 \ -- \ {\bf 00146 ROMA, Italia}}\\
$^2$  {\sf I.N.F.M. - sezione di ROMA.}\\
}
\end{center}
\vskip3.3truecm 
{\large 
{\bf Submitted to}: Planetary and Space Science\\
}
\vskip2truecm
\null\noindent
{\large 
{\bf Send proofs to:} Piero CIPRIANI (address above)
}\\
{\large 
{\bf Send offprint request to:} Piero CIPRIANI (address above)
}\\
\vfill\baselineskip12pt
\noindent
\hrulefill\hspace{9truecm}\ \\
$^\dagger$ {\small{\tt e-mail: CIPRIANI@AMALDI.FIS.UNIROMA3.IT}}\\
$^\ast$    {\small{\tt e-mail: DIBARI@VXRMG9.ICRA.IT}}\\
\end{titlepage}
\Message{Titlepage:}
\setcounter{footnote}{0}
\setcounter{page}{2}

\begin{abstract} 
\baselineskip12pt
In this paper, with particular emphasis on the N-body problem, we focus our
attention on the possibility of a synthetic characterization of the qualitative
properties of generic dynamical systems. The goal of this line of research
is very ambitious, though up to now many of the attempts have been discouraging.
Nevertheless we shouldn't forget the theoretical as well practical relevance
held by a possible successful attempt. Indeed, if only it would be concevaible
to single out a {\sl synthetic} indicator of (in)stability, we will be able
to avoid all the consuming computations needed to {\sl empirically discover}
the nature of a particular orbit, perhaps sensibly different with respect to
another one very {\sl near} to it. As it is known, this search is tightly
related to another {\sl basic question}: the foundations of Classical 
Statistical Mechanics. So, we reconsider some 
basic issues concerning the (argued) relationships existing between the dynamical 
behaviour of many degrees of freedom ({\sf mdf}) Hamiltonian systems and the 
possibility of a statistical description of their macroscopic features. 
The analysis is carried out in the framework of the {\sl geometrical description 
of dynamics} (shortly, {\sl Geometrodynamical approach}, 
{\sf GDA}), which has been shown, \cite{CPTD,Marco93,MTTD,Arg95,PRE97a}, 
to be able to shed lights on some otherwise hidden connections between the 
qualitative behaviour of dynamical systems and the geometric properties of the 
underlying manifold.\  
We show how the geometric 
description of the dynamics allows to get
meaningful insights on the features of the manifold where the motion take 
place and on the relationships existing among these and the stability properties
of the dynamics, which in turn help to interpret the evolutionary 
processes leading to some kind of metastable quasi-equilibrium states.\\ 
Nonetheless, we point out that a careful investigation is needed in order to 
single out the sources of instability on the dynamics and also the conditions 
justifying a statistical description.
Some of the claims existing in the literature have here been reinterpreted, 
and few of them also corrected to a little extent.  
The conclusions of previous studies 
receive here a substantial confirmation, reinforced by means of both a 
deeper analytical examination and numerical simulations suited to the goal. 
These latter confirm well known outcomes of the pioneering investigations 
and result in full agreement with the analytical predictions 
for what concerns the instability of motion in generic classical
 Hamiltonian systems. They give further support to the belief about
the existence of a hierarchy of 
time-scales accompanying the evolution of the system towards a succession 
of ever more detailed local (quasi) equilibrium states. 
These studies confirm the absence of a direct {\it (i.e. trivial)} 
relationship between {\sl average curvature} of the manifold 
(or even frequency of occurrence of negative values) 
and dynamical instability. Moreover they clearly single out 
the full responsibility of fluctuations in geometrical quantities in driving a 
system towards Chaos, whose onset do not show any correlation with the average
values of the curvatures, being these almost always positive. These results
highlight a strong dependence on the interaction potential of the 
connections between qualitative dynamical behaviour and geometric features. 
These latter lead also to intriguing hints on the 
statistical properties of generic many dimensional hamiltonian systems.
It is also presented a simple analytical derivation of the scaling law 
behaviours, with respect to the global parameters of the system, 
as the specific energy or the number of degrees of freedom, of the relevant
{\sl geometrodynamical} quantities entering the determination of the
stability properties.
From these estimates, it emerges once more the strong peculiarity of 
Newtonian gravitational interaction with respect 
to all the topics here addressed, namely, the absence of any threshold, in 
energy or number of particles (insofar N$\gg 1$) or whatsoever parameter, 
distinguishing among different {\sl regimes} of Chaos; the peculiar character 
of most geometrical quantities entering in the determination of dynamical 
instability, which is clearly related to the previous point and that it is 
also enlightening on the physical sources of the statistical features of 
N-body self gravitating systems, contrasted with those whose potential 
is {\sl stable} and {\sl tempered}, in particular with respect to the issue 
of {\sl ergodicity time}.
We will also address another {\sl elementary} issue, related
to a careful analysis of {\sl real} gravitational N-body systems, 
which nevertheless provides the opportunity to gain relevant 
conceptual improvements, able to cope with the peculiarities 
mentioned above.
\end{abstract}
\newpage
\baselineskip13pt
\section*{Introduction}

The discipline of Deterministic Chaos, in the last decades, has known an 
expanding phase hardly comparable with other fields of mathematical physics. 
The increased evidence of the ubiquity of instability has spurred an improved 
understanding of nonlinear dynamics.
However, at present, as usual for quickly expanding areas of research, after 
the first phase of {\sl spontaneous growth}, the demand for a more systematic 
settlement, together with a knowledge of the chaotic phenomena going beyond 
the one of semi-phenomenological nature, is also rising.\ 
The signature of Chaos in realistic models of physical systems is indeed often 
phenomenological in character, as the occurrence of instability is usually 
recognized {\it a posteriori}, looking at its consequences on 
{\sl unpredictability} through the study of the evolution of a generic 
perturbation (which represents, from a physical viewpoint, the unavoidable 
uncertainty on the initial conditions) to a given reference trajectory. 
However, recently there has been a renewed interest in looking
for, at least qualitative, synthetic indicators of dynamical instability in 
realistic models of physical systems, able to provide an {\it a priori} 
criterion. As far as we know, the first {\sl explicit} concrete attempt in 
this direction dates back to Toda's paper, \cite{Toda}, whose goal was 
to find a concise characterization of the sources of {\sl global} 
instability related to {\sl local} criteria.
As it has been shown (see,\eg, \cite{criticaToda,Casati}), the 
relationship argued by Toda was somewhat too {\it na\"\i ve}, 
and the
link between local and global instabilities turns out to be by far more 
complex than sought.\\
Nevertheless, mainly in the field of celestial mechanics, where the 
dynamical systems under study possess few degrees of freedom ({\it few}, 
here, means of order ten, or less), a number of attempts to single out synthetic 
indicators of instability (although qualitative), have been put forward, 
looking at the intermingled structure of deformed or destroyed {\sl invariant 
surfaces} in phase space. This effort has led to a deeper 
understanding of the conditions responsible for the onset of Chaos in 
dynamical systems of low dimensionality and, in a sense, not too far 
from integrable ones.\\
At the other end, very {\sl far} from integrability, within the sphere of Ergodic 
Theory, with reference to Statistical Mechanics
(although not so close to the latter as the very origin and motivations of 
the former should let to expect), and consequently mainly for \mdf systems, 
some results have been obtained in the case of abstract dynamical systems, 
\cite{Anosov,Sinai}, but they seem to be of little utility to gain some 
insights on the behaviour of realistic physical models.\ Whereas, being 
mainly focused on conceptual issues on the basic postulates of Analytical
and Statistical 
Mechanics, and for this reason relevant to the present work, the last 
decades knew a deeper reconsideration of the issue about the origin of
unpredictability and of the problem 
related to the approach to equilibrium for generic \mdf Hamiltonian systems 
(see, \eg, \cite{Galgani} and references therein). This {\sl problem} 
arose after the first {\sl numerical experiments} performed on a computer, 
when Fermi, Pasta and Ulam, \cite{Fermi}, tried to follow numerically the dynamics 
of a chain of weakly coupled nonlinear oscillators, in the hope of
observing the theoretically expected process of equipartition of 
energy among normal modes. 
Those {\sl experiments}, because of their {\sl surprising} results, stimulated a 
lively debate, and supported the believe that the semi-heuristic 
argumentations of Boltzmann himself and Jeans about the {\sl exponentially long 
equipartition times} amongst high and low frequency modes possess a 
very deep physical meaning\footnote{For a very interesting 
account on these pioneering  and, for a long time, rather disregarded intuitions,
see again \cite{Galgani}.}.\\
Among the attempts towards intrinsic (or synthetic) criteria to single out 
the presece of Chaos in models of concrete physical systems, besides the 
Toda's trial already mentioned, we recall the Krylov's Ph.D. thesis,
\cite{Krylov}, whose intuitions had important influence on the whole line 
of research this work belongs to.\\
Nevertheless, the founding arguments of the investigation of the stability 
of motions by means of a geometrization of the dynamics date back to
the paper by Synge, \cite{Synge}, which also partially summarizes previous works
of Ricci and Levi-Civita. Although completely unaware of the 
relevance of exponential instability (of course, in 1926!), he there gave a 
very clear exposition of the power of the {\sl Geometrodynamics} in gaining 
intrinsic information on the qualitative behaviour of generic holonomic 
dynamical systems. Other authors, \cite{Eisenhart}, since then, 
added their own contribution to the development of the subject, extending the 
range of applicability of the method. In the recent years a new deal of 
interest has been raised, stimulated by the Krylov's work cited above, 
(unfortunately diffused worldwide more than two decades after its appearance) 
and from the mess of papers produced by the abstract ergodic theory (see, 
\eg, \cite{Sinai}). But only very recently the approach has received 
a self-consistent settlement able to cope with the occurrence of instability 
in \mdf hamiltonian systems (see \cite{Marco93,CPTD}) and to give a key to 
understand the links between local properties of the manifold and the asymptotic 
qualitative features of the dynamics.\ In a recent paper, \cite{PRE97a}, we 
extended further the {\sl Geometrodynamical approach}, to handle 
a wider class of dynamical systems, even with peculiar lagrangian structure, 
embedding the dynamics in an ensemble of manifolds more general than riemannian: 
the Finsler ones. This framework enable us to cope with dynamical systems 
otherwise not suitable for a geometrization within the riemannian geometry,
\cite{Sigrav96}, to get more insights on the nature of instability,
to locate the relationships with the geometrical structure of the manifold,
\cite{CPMT_HH}, and to set up a formalism with a
{\sl built-in} explicit invariance with respect to any reparametrization of 
the time variable (see also \cite{Aquila2}).\ The {\sf GDA} has proved
to give many interesting insights on the analysis of qualitative properties
of dynamical systems ({\sf DS}) of interest in various fields of Statistical 
Mechanics, as well in Celestial Mechanics and Cosmology; 
we will address here a problem that,
also looked from a {\sl number criterion}, seats just in the middle between
Analytical ($\log{N}\sim 1$) and Statistical ($\log{N}\sim 23$) Mechanics. 
The main concern of the present paper is indeed towards a clarification of the 
links between geometry and dynamics on one side, and their implications on the 
statistical behaviour of N-body systems, when $6\minord\log{N}\minord 13$, 
on the other.
In the sequel, we will focus on the properties of the gravitational N-body system 
and, using the Fermi-Pasta-Ulam ({\sf FPU}) chain as a {\sl normal 
reference system}, belonging to the class of \mdf systems for which a 
Statistical Mechanical description is well founded,
compares their dynamical, geometrical and statistical features, 
enlightening from a perhaps nouvel perspective the physical and mathematical 
peculiarities of Gravity, showing also how to treat successfully some of them.

\subsection*{Instability and Statistical description.}
The relevance of dynamical instability and ergodicity\footnote{We do not
intend here neither to confuse nor to assimilate the two properties.
We would only point out how even some very basic questions concerning the 
links between Analytical and Statistical Mechanics are still awaiting for an
universally accepted, although not rigorously proved, answer.} 
to Statistical Mechanics ({\sf SM}) has been sometimes questioned, 
\cite{Farquhar}, but it is by now generally accepted that the 
unpredictability implied by Chaos is a natural ingredient to 
justify the statistical description of \mdf dynamical systems. As stressed in 
the introduction, however, there is no rigorous, or even convincing, prove of 
any direct relationship between the instability times, as can be derived from 
dynamics (Lyapunov exponents, Kolomogorov-Sinai entropy,...) and the 
time-scales related to a Statistical Mechanical treatment, \ie, those belonging 
under the name of {\sl relaxation times}. Nevertheless some qualitative 
relationships exist among them for some interesting and historical \mdf 
systems, as the {\sf FPU} chain, for which, although differing each other 
by orders of magnitude, the Lyapunov and the \SM relaxation times display the 
same transition between two different regimes of {\sl weak} and {\sl strong} 
chaoticity, \cite{Marco91}.\ 
Recently (see \cite{Froeschle} and references therein), also for dynamical 
systems with few degrees of freedom and of particular interest in Celestial 
Mechanics, some debate raised about such a relation among Lyapunov exponents and
the time scale of {\sl diffusion} in phase space.\\
A thorough treatment of the multiple connection between dynamical instability 
and Statistical Mechanics is out of place here, \cite{CPTD,PRE97b}. 
Instead, we would like to stress that the suggestions of Boltzmann 
and Jeans, reminded in the previous section, received, in the framework of 
Analytical Dynamics, a strong support from the Nekoroshev theorem on the 
exponentially long time scales of {\sl quasi-conservation} of actions for 
generic non integrable systems, \cite{Nekoroshev}. This is relevant to the 
present discussion in giving an explanation to the existence of different 
regimes of stochasticity in generic Hamiltonian systems of interest in 
Solid State Physics as well in Astrophysics, and also in demonstrating 
once more the peculiarity of Newtonian Gravity. 
Indeed, for the N-body system governed by gravitational interaction, we
will support the claim according to which simply do not exist a regime
of {\sl quasi integrability}, as this system will be {\sl chaotic} for almost 
all generic initial conditions\footnote{We are supposing obviously that 
$N\gg 1$, and we speak about {\sl chaos} in a not too strict sense, as the 
long range behaviour of the interaction, more than its short range divergencies, 
causes the phase and configuration spaces of the system to be non compact.}. 
The absence of any integrable limit, makes it hard to find out any threshold,
$\varepsilon_*$, for the magnitude of the perturbation, $\varepsilon$, below 
which the actions are {\sl quasi conserved} for time intervals exponentially 
long in some power of $(\varepsilon_*/\varepsilon)$; simply, we don't know
the state for which $\varepsilon=0$!

  Behind the long standing problem of the justification of the {\sl basic 
assumptions} at the ground of the Statistical description, \cite{Ma}, for 
generic dynamical systems, lies a slowly evolving set of ideas, 
all referred to as the {\sl Ergodic hypotesis}\footnote{The 
{\sl ensemble}, often very inhomogeneus, of concepts indicated under this name 
has sometimes been presented in a somewhat misleading fashion, leading, 
from time to time, either to evidently unreliable assumptions or to trivial 
assertions. This has been an undeserved treatment to Boltzmann efforts to
find a strict formulation for a very deep and notwithstanding {\sl ineffable}
physical intuition.
 A thorough and enlightening conceptual re-examination of the issues related to 
the Boltzmann ideas on the importance of a {\sl physically reasonable 
formulation} of the Ergodic hypotesis to 
justify the use of Statistical approach in Mechanics has been carried out 
recently in a series of paper by G.Gallavotti; for an exhaustive account, see 
the monography \cite{GallaMS}.}.
The {\sf GDA}, applied to N-body gravitational 
system, discloses evidence of the differences amongst some of the 
interpretations given of it in order to justify the approach to equilibrium, 
and, at the same time, reinforces the claim that the processes leading a 
{\sf DS} to the hierarchy of 
intermediate partial equilibria, proceed on time scales whose spectrum is {\sl 
written} in the interaction potential governing its dynamics, and can differ by 
orders of magnitude passing from a \DS~ to another. We refer to 
\cite{CPTD,PRE97b} for a less synthetic discussion of these issues, and 
concentrate here mostly on the dynamics of N-body systems and on the insights 
on their qualitative properties that can be obtained in the framework of \GDA.

\section*{Geometrodynamics of evolutionary processes.}
The problem of the approach to equilibrium in physical systems with a large
number of degrees of freedom is, essentially, a problem of time-scales. The
fundamental question is: what time-scale is required such that the system finds
itself in a state in which some or all memory of the initial one is lost?  
A theory of non-equilibrium processes that should be able to give quantitative 
answers
to this question would also provide the hints to solve most of the open
issues referring to the connection between Analytical and Statistical
Mechanics. This problem has been investigated since the very birth of {\sf SM}
with different methods; here we will exploit the Geometrical approach.\\
 In the following sections we recall briefly the method 
and then introduce the main Geometrodynamical Indicators 
({\sf GDI}'s), whose behaviour determine the qualitative (global) evolution 
of large N-body systems.\\
In order to highlight the singular features of Gravity, we apply the method 
to two potentials, representative of the classes of short and long range 
interactions respectively.  
The evaluation of the characteristic time-scales for them is firstly derived 
on the basis of analytical estimates of the quantities entering the {\sf GDI}'s. 
On these grounds some of the existing results are reviewed and 
reinterpreted, giving an (hopefully) coherent scenario. After that, the 
reliability of the claims based on a trivial extension of methods and concepts 
borrowed from Ergodic Theory are critically examined and discussed. 
Still within a semi-analytical approach, we show why that extension results, 
at least, too much na\"\i ve. Using numerical simulations suitably 
tailored to the goal, we then show that 
the analytical estimates correctly describe the behaviour of {\sf GDI}'s, 
and so that the mechanisms driving to Chaos in \mdf Hamiltonian systems 
differ completely from those occurring within the Ergodic theory of abstract 
\DS's.\\
We conclude comparing the behaviour of the {\sf GDI}'s for the {\sl mathematical} 
Newtonian potential, on a side, with that of the corresponding quantities
related to the {\sl physical} gravitational N-body problem and to the 
\FPU chain, on the other. This allows us to shed some lights on the way a 
stellar systems could attain its apparent equilibrium state, in spite of 
the celebrated theorems of {\sl rigorous} Statistical Mechanics on 
non-existence of a {\sl maximum entropy state} for unscreened Coulombic 
interactions, which do not possess the property of {\sl stability},
\cite{Ruelle}. 
The search for the conditions needed in order to guarantee the equivalence
between {\it time} and {\it phase} averages of geometric quantities, leads also 
to the conclusion that the peculiarities of Newtonian gravity, reflects 
themselves in a singularity on the {\sl literal ergodicity time}, singularity 
which disappears when a realistic interaction, that nevertheless do not modify
the very nature of Gravity, is considered, irrespective to the details of
the {\sl care}.\\
Though there are several {\sl geometrizations} of Dynamics, 
in the following, we exploit mostly that based on the 
so-called {\sl Jacobi} metric, 
which results from a straightforward application 
of the {\sl least action principle} in the form given by Maupertuis, 
see, \eg, \cite{Goldstein,ArnoldMC}.

\subsection*{The Jacobi Geometrodynamics.}
In the framework of a definitely non-perturbative treatment of the Hamiltonian
Chaos, the differential geometrical approach stands in the pathway that,
started by Krylov with the aim to comprehend the relaxation
processes in realistic physical systems, has grown very long in the
applications to abstract dynamical systems, acquiring mathematical
strictness but only recently broadening the physical interest. The Geometrodynamical 
approach to a Hamiltonian system reduces
it to a geodesic flow over a manifold, {\sf M}, on which is defined a suitable 
metric. 
In the case of the Jacobi geometrization, the manifold is a riemannian one, 
equipped with the (conformally flat) metric, ${\sf g} = \{g_{ab}\}$,
whose line element is
\be
ds_J^{\ 2}\ \defin g_{ab} dq^a dq^b\ =\ \left[E - {\cal U}(\bf q) \right] 
\eta_{ab} dq^a dq^b = 2\,\Wcal^2\,dt^2,
       \qquad a,b = 1, \dots, {\cal N}\ ;
\label{metrica} 
\ee
where $E$ is the total conserved energy, ${\cal U}({\bf q})$ is the potential 
energy, depending on the coordinates $\q=\{ q^a\}$ on the {\it configuration 
manifold} of the system (with ${\cal N}$ degrees of freedom), $\Wcal (\q ) 
\defin [E-\Ucal (\q )]$ is the {\sl conformal factor}, whose magnitude
numerically equals given the kinetic energy ${\cal T}$, and $t$ is the 
newtonian time. 
The metric of the $\Ncal$-dimensional
{\it physical space} of the system is thus
$\eta_{ab}$, \ie, $2{\cal T} = \eta_{ab} \dot{q}^a \dot{q}^b$,
 which reduces to $\eta_{ab}\equiv\delta_{ab}$ when the space 
is euclidean and cartesian coordinates are employed; as usual, we denote 
with a dot differentiation with respect to time. Since we are considering 
N-body Hamiltonian systems, we have $ {\cal N}
\equiv {\rm N} \cdot d $, where $d$ is the dimensionality of the system. 
So, for the self-gravitating system we have $d=3$, whereas $d=1$ for the 
\FPU chain.\\
In the language of \GDA, the ${\cal N}$ second order Lagrangian equations 
of motion
(or the $2\Ncal$ first order Hamiltonian ones) are replaced by the $\Ncal$ 
geodesics equations
\be
{\displaystyle {{\nabla u^a}\over{ds}}} \equiv 
{\displaystyle {{d u^a}\over{ds}}} + \Gabc u^b u^c =0\ , \qquad
(a=1,\ldots,\Ncal)
\ee
where $u^a = dq^a/ds$, $\nabla/ds$ is the total (covariant) derivative 
along the flow, the $\Gabc$ are the Christoffel symbols of the metric
${\sf g}$ and the summation convention is understood.\\
 For the metric defined in (\ref{metrica}), namely $g_{ab} = \Wcal (\q ) 
\eta_{ab}$, and using cartesian coordinates, the geodesic equations 
read\footnote{Since we restrict to 
consider mostly Jacobi geometrization, we will from now on omit the subscript in 
$ds_J$ whenever there is no risk of ambiguity, writing simply $ds$.}:
\be
{\displaystyle { {{d^2q^a}\over {ds^2}} + {1\over {2\Wcal}} 
\left[ 2{{\partial \Wcal}
\over {\partial q^c}} {{dq^c}\over {ds}}{{dq^a}\over {ds}} -
g^{ac} {{\partial \Wcal}\over {\partial q^c}} \right] } =0, }\ , \qquad
(a=1,\ldots,\Ncal) 
\label{EG1} 
\ee
and give the trajectories on the configuration manifold in terms of the affine
parameter $s$ (conciding with the {\it Maupertuis action}). 
The {\sl newtonian time} law of percurrence of the trajectory is obtained via 
the relation $ds=\sqrt{2}\, \Wcal(\q )\, dt $, exploiting which one obtain the 
familiar equations of motion, $\ddot{q}^a + \Ucal^a = 0$.\\
Once rephrased the dynamics as a geodesic flow, we are left with the 
determination of the question of stability of geodesic paths on the Jacobi 
manifold. Within the framework of Hamiltonian dynamics this issue is 
addressed using the tangent dynamics equations, which determine the evolution 
of a small deviation vector, $\D=(\bfxi,\dot\bfxi)$, in phase-space, describing a 
perturbation to a given trajectory.\\
Although the equations for the trajectories coincide with geodesics ones
once rephrased in terms of the same parameter,
the equations for the {\sl displacements}, which are those relevant to the
issue of stability of motion, differ in general.  We refer to \cite{Aquila2} 
for a discussion of the problem of {\sl equivalence} among different 
{\sl linearized} equations for displacements and here simply remind that 
all geodesics have, by definition, unit velocity vector, so, in a sense, 
the deviation between geodesics gives the distance between {\sl paths} and not 
between points on them. Another point to be remarked concerns the apparent self 
consistency of the geometrical description, which allow for a built-in 
{\sl distance}, lacking in the
usual definition for the norm of a vector in phase space, where  
an Euclidean structure is imposed by hand.

\section*{Geometric description of instability.}
As it is well known, in the framework of Riemannian geometry\footnote{And 
also in more general differential manifolds, see \cite{MTTD,PRE97a}.} the 
evolution of a perturbation, $\delta\q$, to a geodesic is described by the 
Jacobi--Levi-Civita ({\sf JLC}) equation for {\it geodesic spread}: 
\be
{{\nabla} \over {ds}}\left({{\nabla \delta q^a} \over {ds}}\right)
 + {\cal H}^a{}_c \delta q^c = 0, 
\label{EDG}
\ee
where it has been defined the
{\sl stability  tensor} ${\cal H}^a{}_c$, defined along any geodesic
and related to the Riemann curvature
tensor, $R^a{}_{bcd}$, associated to $ g_{ab} $ by
\be
{\cal H}^a{}_c \defin  R^a{}_{bcd} u^b u^d\ ,
\ee
where $\u$ is the unit tangent vector to the geodesic. In the case of Jacobi 
metric \bfHcal~ reads
\begin{eqnarray}
{\cal H}^a{}_c & = & 
         \left[ {1\over {4\Wcal^2}} \left( {d\Wcal}\over {ds} \right) ^2 -
          {1\over {2\Wcal}} {{d^2\Wcal}\over {ds^2}} \right] \delta^a_c +
	   {{( \grad\Wcal )^2}\over {4\Wcal^3}} u^au_c + 
\nonumber \\
      &  & + {1\over {2W}} \left[\Wcal,_{bc} (u^a u^b - g^{ab}) 
           + g^{ab} \Wcal,_{bd} u^d u_c  \right] + \\
      &  & - {3\over {4\Wcal^2}} \left[
          \left( {d\Wcal}\over {ds} \right) (u^a\Wcal,_c + g^{ab} 
            \Wcal,_b u_c )- g^{ab}\Wcal,_b\Wcal,_c \right] \nonumber 
\label{Htens}
\end{eqnarray}
which clearly depends also on the position, $\q\in {\sf M}$.\ Here, as usual,
$\Wcal,_a=\partial\Wcal/\partial q^a$, the summation convention is adopted,
and the Euclidean $\Ncal$-dimensional gradient operator acting on a function
$f$ defined on {\sf M} has been indicated explicitly as $[\grad f]$, rather
than $\nabla f$, to avoid 
confusion with the already defined total derivative along the flow $\nabla/ds$.

Equation (\ref{EDG}) contains all the information about the evolution of a 
congruence of  geodesics emanating within an initial distance 
$z = \| \delta\q\|=({g_{ab}~\delta q^a \delta q^b})^{1/2}$
from the reference one. The problem for a system with \mdf is that the 
Riemann tensor contains ${\cal O} (\Ncal^4)$ components, and the evaluation 
of the stability tensor \bfHcal~ is an huge task. Nevertheless, just when the 
dimensionality of the ambient space is high, some assumptions can be made 
on its {\sl global} properties, averaging in a suitable way. This procedure 
has been adopted in the past (see, \eg, \cite{Marco93,CPTD} and references 
therein) with success in describing the behaviour of \mdf systems of interest 
in Statistical Mechanics.
Indeed, if we are interested in the stability properties of the flow, the 
relevant quantity is the magnitude of the perturbation, measured within the 
\GDA~by its norm. Indeed, 
if $\delta\q=z\bfnu$, where \bfnu~is a vector of the unitary
tangent space at $\q$, ${\sf T}_{\q}{\sf M}$, 
the norm $z$ evolves according to:
\be
{{d^2z}\over{ds^2}} = \left( -\Hcal_{ac} \nu^a \nu^c + \s| 
{{\nabla\bfnu}\over{ds}}\d|^2 \right) z 
\label{norma1}
\ee
which is still an {\sl exact} equation derived directly from eq.(\ref{EDG}), 
\cite{CPTD}. However, eq.(\ref{norma1}) 
contains still the full stability tensor, and the task has so been made
easier only partially, reducing to a single one the number of equations 
to be integrated. To {\sl close} this equation some 
assumption is needed.

\subsection*{Geometric Indicators of Instability.}
We refer to \cite{CPTD,Marco93} and subsequent works for the description 
of the details of the averaging procedure, and we want here recall only 
some critical points which deserve more cautious treatment than the one 
given sometimes, \cite{GS,Kandrup}. Indeed, one intuitive approximation that 
help to partially get rid of the huge amount of computational task in the 
evaluation of the stability tensor $\bfHcal$ is suggested by the physical 
reasoning that we are interested in the evolution of a {\sl generic} 
small perturbation, however oriented with respect to the {\sl given geodesic}. 
Actually, the $\bfHcal$ tensor contains all the informations about the 
{\it local} evolution of {\it any} perturbation to the system, {\it local} 
in the sense that it describes the behaviour of the vectors of tangent 
space in a neighbourhood of the {\sl state} $(\q,\u)$. 
So, a first averaged equation can be derived from eq.(\ref{norma1}) 
letting $\delta\q$ to have a random orientation {\sl with respect 
to the {\it given} flow}. 
So, as we have by definition that 
$\|\bfnu\|^2\equiv g_{ab} \nu^a\nu^b =1$,
being the Jacobi metric conformally flat, this equation would be 
consistent with a randomly chosen orientation for \bfnu~if we put
$ \langle \nu^a\nu^b\rangle = g^{ab}/\Ncal $.
Nevertheless, while practically irrelevant when $N\gg 1$, we
should make here a comment about the correct way of averaging the deviation
over directions. Obviously, given a geodesic, there are $(\Ncal-1)$ independent
orientations orthogonal to \u, and we can imagine also a perturbation
with a non vanishing component along the geodesic itself. 
This straight argumentation leads to the $\Ncal^{-1}$ factor in the average above.
Yet, it follows directly from the very definition of the
{\sl stability tensor}, \bfHcal, \ie, from the simmetry properties of 
Riemann tensor, that at least one amongst its local eigenvalues 
vanishes, being that associated with the direction along the flow:
\be
\Hcal^a{}_c~ l^c \equiv 0,\ \ \qquad\ \forall\ \ {\bf l} = P \u ,
\label{pupa}
\ee
with $P(s)$ any scalar quantity.
So, a deviation along the geodesic cannot evolve more than linearly in
the $s$-parameter, and consequently do not contribute to exponential instability.
Moreover, it is known, \eg, \cite[\S 17.6]{Katok}, that the evolution of
any parallel component $z_{{}_\parallel}$ decouples from that of the remaining
{\sl normal} one $z_{{}_\perp}$; then, as interested in the possibly
exponential growth of the deviation, we are left with only $\Ncal-1$ independent
orientations of \bfnu, once fixed the geodesic. Said in other words, we consider
only perturbations orthogonal to the geodesic, so that
$$
g_{ab} \nu^a \nu^b =1\qquad ;\qquad g_{ab} \nu^a u^b =0\ ,
$$
which leads to
\be
\langle \nu^a\nu^b\rangle = {{g^{ab}}\over{\Ncal-1}}\ .
\ee
Within this {\sl reasonable and self consistent approximation}, equation 
(\ref{norma1}) reduces to the averaged one:
\be
{{d^2z}\over{ds^2}} + k_\u (\q)\, z = 0
\label{norma2} 
\ee
where 
\be
k_\u = k_\u (\q) \defin {{1}\over{\Ncal-1}} \Hcal^a{}_a = {{1}\over{\Ncal-1}} 
{\sf Tr}\,\bfHcal\ =\ {{1}\over{\Ncal-1}} {\Sum_{A=1}^{\Ncal-1}} 
K^{(2)} (\u,{\bf e}_A)\ ;
\label{curv}
\ee
from now on, as above, the dependence on $\q\in{\sf M}$ of all geometric
quantities will be understood whenever no confusion can arise.\\
We now sketch the line along which it can be shown, \cite{CPTD}, that this 
quantity equals the (normalized) {\sl Ricci 
curvature} in the direction of the flow:
\be
k_R (\u ) \defin {{\ric (\u)}\over{\Ncal-1}}\ ,
\ee
being the Ricci curvature along any direction indicated by the
unit vector ${\bf e}_A\in {\sf T}_\q{\sf M}$, defined as
\be
\ric ({\bf e}_{{}_A}) \defin  R_{bc} e_A^b e_A^c\ ,
\label{ric_A}
\ee 
where $R_{ac}\defin R^b{}_{abc}\equiv g^{bd}R_{abcd}$ is the Ricci tensor.\\
The quantities $K^{(2)} (\u,{\bf e}_A)$ in eq.(\ref{curv}) are the sectional 
curvatures in the $(\Ncal-1)$ independent planes spanned by $\u$ and the 
$\{ {\bf e}_A\},\ (A=1,\ldots,\Ncal-1),\ ({\bf e}_A,{\bf e}_B)=\delta_{AB},\ 
({\bf e}_A,\u)=0$. 
In general, given a riemannian manifold, $({\sf M, g})$, the definition of 
sectional curvature 
in the plane spanned by two non parallel vectors, $(\x,\y)\in \TqM$ is defined as
\be
K^{(2)} (\x,\y ) \defin {\displaystyle{{R_{abcd} x^a y^b x^c y^d}\over 
{\|\x\|^2~\|\y\|^2 -
\|\x\cdot\y\|^2}}} \equiv K^{(2)} (\y,\x )\ ;
\label{sectcurv}
\ee
where the norm and the scalar product are those induced by {\sf g}.\\
Without loss of 
generality, \cite{SyngeSchild}, we can suppose $\x\perp\y$; so, 
given an $\Ncal$-dimensional 
manifold, by simmetry it follows that
there are $\Ncal(\Ncal-1)/2$ independent sectional curvatures. Further, 
given an orthonormal
basis in {\sf M}, $\{ {\bf e}_{{}_A}\},\ (A=0,\ldots,\Ncal-1)$, we indicate 
these sectional curvatures as
\be
K^{(2)}_{{}_{AB}} \defin K^{(2)} ({\bf e}_{{}_A},{\bf e}_{{}_B}) 
\equiv R_{abcd}~ e_{{}_A}^a
e_{{}_B}^b e_{{}_A}^c e_{{}_B}^d \equiv K^{(2)}_{{}_{BA}},\ \ (B\neq A).
\ee
We preliminarly show that the {\sl Ricci curvature} along the direction 
${\bf e}_{{}_A}$ defined above, eq.(\ref{ric_A}) is also equal to
\be
{\sf Tr}\RcA = \RcA_{bd}{\Sum_{B=0}^{\Ncal-1}}{}^{'} e_{{}_B}^b  e_{{}_B}^d
\label{Tr_RcA}
\ee
where we defined the $\Ncal$ tensors $\RcA_{bd}\defin R_{abcd}~ e_{{}_A}^a 
e_{{}_A}^c$ and the prime stands to indicate that the sum is over $B\neq A$.
Indeed, in order to see that the \rhs of eq.(\ref{Tr_RcA})
actually equals the trace of the corresponding tensor, we observe that, by
definition,
$$
\RcA_{bd}{\Sum_{B=0}^{\Ncal-1}}{}^{'} e_{{}_B}^b  e_{{}_B}^d = 
R_{abcd}~ e_{{}_A}^a e_{{}_A}^c {\Sum_{B=0}^{\Ncal-1}}{}^{'} 
e_{{}_B}^b  e_{{}_B}^d\ ;
$$
and, using the anti-simmetry properties of the Riemann tensor, we can include in
the summation also the term $B=A$, which vanishes identically.\\
Being $\{ {\bf e}_{{}_A}\}$ an orthonormal basis, we then have
\be
\Sum_{B=0}^{\Ncal-1} e_{{}_B}^b  e_{{}_B}^d = g^{bd}\ ,
\ee
so that:
\be
R_{abcd}~ e_{{}_A}^a e_{{}_A}^c {\Sum_{B=0}^{\Ncal-1}} 
e_{{}_B}^b  e_{{}_B}^d =
\RcA_{bd}~ g^{bd} \equiv \RcA^b{}_b = {\sf Tr}
\bfRcA\ = g^{bd} R_{abcd}~ e_{{}_A}^a e_{{}_A}^c = R_{ac}~ e_{{}_A}^a e_{{}_A}^c
=\ric ({\bf e}_A)\ ;
\label{ric=trace}
\ee
in such a way that the Ricci curvature along any vector
of the unitary tangent space coincides with the trace of the
corresponding $\bfRcal$ tensor. 
Incidentally, from the leftmost hand side of eq.(\ref{ric=trace}) we
see
\be
\ric({\bf e}_A) = {\Sum_{B=0}^{\Ncal-1}}{}^{'} K^{(2)}_{{}_{AB}}\ ;
\ee
that is enough to justify also the last equality in eq.(\ref{curv}).
Now, given a geodesic passing through $\q\in {\sf M}$, let $\u\in\TqM$ its 
unitary tangent vector. If we put ${\bf e}_{{}_0}=\u$, we see that what we
called {\sl stability tensor} for the given geodesic is actually
\be
\Hcal^a{}_b \equiv \Rc0^a{}_b\ ,
\ee
and this completes the proof that 
${\sf Tr} \bfHcal \equiv \ric(\u)$, \ie, 
$k_\u = k_R(\u)$.\\
As we have just seen, to define the $(\Ncal-1)$ sectional curvatures when 
one direction, \u, has been fixed as the tangent vector to a given geodesic it 
suffices to choose the remaining $(\Ncal-1)$ orthogonal unit vectors 
$\{ {\bf e}_{{}_A}\},\ (A=1,\ldots,\Ncal-1)$, transversal to the flow. 
A preferred set is however obtained if we furthermore introduce the 
{\it principal sectional curvatures} for the congruence defined by $\u$, 
\ie, the {\sl local} eigenvalues, $\{\lambda_{{}_A}\}$, of the stability tensor  
such that
\be
\Hcal^a{}_c\ {\bf e}_{{}_A}{}^c = \lambda_{{}_A} {\bf e}_{{}_A}{}^a,\qquad 
(A=0,\ldots,\Ncal-1)
\label{lambdak}
\ee
with normalized eigenvectors, $\|{\bf e}_{{}_A}\|=1$\ ;
so, by eq.(\ref{pupa}), it turns out 
\be
\lambda_{{}_0} = 0\ \ \ \ \ {\rm if}\ \ \ \ \ {\bf e}_{{}_0} = \u\ .
\ee
and also that the Ricci curvature per degree of freedom $k_R(\u )$ 
represents the average of the non trivial principal sectional curvatures 
$\{\lambda_{{}_A}\}$\footnote{As we observed, from eq.(\ref{pupa}) it follows 
that the non trivial geodesic deviations should have a component orthogonal 
to the flow, as a parallel perturbation cannot contribute to the possible 
instability. 
Stated otherwise, the knowledge {\it a priori} of a vanishing eigenvalue 
of $\bfHcal$ suggests to divide the trace by $(\Ncal-1)$ instead than $\Ncal$; 
this is also consistent with the fact that, given a direction, there are only 
$(\Ncal-1)$ independent 2-planes containing it. 
As we were interested in \mdf systems, we wouldn't need here to pursue further 
this point, however taking into account such a distinction below, when we 
average also on the orientations of $\u$, counting correctly the number 
of (ordered) direction pairs as $\Ncal(\Ncal-1)$; also because it has to be
remarked the great importance held by this factor in the applications of
the \GDA~to few dimensional \DS's.\label{notaN}}.\ 
From this last definition, it follows in a perhaps
more direct way the well known result, \eg,\cite{SyngeSchild}, according 
to which the sum of all the sectional curvatures in the $(\Ncal-1)$ 2-planes
containing a fixed direction do not depends on the choice of the set of
independent normal directions, being simply the trace of the tensor
\bfRcA.
In principle, the $(\Ncal-1)$ non trivial eigenvalues of \bfHcal~form a set 
of indipendent {\it local} indicators of stability, in what they approximately 
determine the {\it local} behaviour of a deviation vector along the 
corresponding eigendirection; indeed, the analysis carried out for few 
dimensional systems, \cite{CPMT_HH}, has shown that the connection between 
the sectional curvatures 
and the dynamical behavior do exist, although it emerges  completely 
when considered 
globally, and that estimates based only on {\sl na\"\i ve} local analysis can 
lead only to partial answers.\ It has been claimed that there is a trivial 
counterexample to this, represented by the flows on manifolds of constant 
curvature. But it isn't at all a counterexample, as, in that case, {\sl 
Schur's theorem}, \cite{SyngeSchild}, assures that there is no distinction 
between {\sl local} and {\sl global} features of the 
manifold\footnote{We will often 
return in the sequel to some underestimate implications of this 
theorem, which results of fundamental importance to shed light on the 
properties of the manifold really tied up to the onset of Chaos.}. 
So, although it is generally true that the qualitative behaviour of geodesics 
depends on global rather than local features (and this is the reason of the 
general failures of {\sl Toda-like} criteria) within the \GDA~we feel
the perception of the intermingled relationships amongst them, which
allows to gain some {\sl global} informations from a {\sl local} analysis 
of the curvature proprties of the manifold.

\subsection*{Curvature, Instability and Statistical Behaviour.}
It has repeatedly been claimed that the chaotic {\sl and} statistical 
properties of dynamical systems, in the case of \mdf, \eg,\cite{GS,Kandrup}, 
and for few dimensional systems too, \cite{Szydlowsky}, depend on the 
{\sl scalar curvature} of the manifold. 
Instead, it has been shown with concrete applications, see, for example,
\cite{Marco93} or \cite{CPTD}, that 
this quantity gives no information at all on the behaviour of trajectories 
for \mdf systems, moreover, using the Eisenhart geometrization, it is found
to vanish identically, resulting  in general a very scarcely reliable
indicator. The reasons of this failure in both few and many dimensional \DS's
turn out easily if we think over the assumptions implicitly or explicitly made
to justify its use.\\
The use of scalar curvature has been accounted for neglecting the 
orientation of the velocity
$\u$, so averaging on {\sl different states} of the system, in addition 
to different orientations of deviation. 
In \cite{CPTD} the consequences of such an approximation 
are analysed thoroughly; here we show the formal derivation and critically 
analyse what would be its implications.\\
If also the tangent vector to the geodesic $\u$ is oriented randomly, 
as $\|\u\|\equiv 1$ too, we can write, in analogy with what has been done 
above for \bfnu, except that now we really have $\Ncal$ independent orientations
(and then $\Ncal-1$ for the orthogonal deviation vector):
\be
\langle u^a u^b \rangle = {\displaystyle{{g^{ab}}\over {\Ncal}}}
\ee
With this strong assumption, we see that the stability tensor loses its 
dependence on the {\sl actual state}, forgetting the information about 
the orientation of the geodesic, and the first term in the \rhs~of 
eq.(\ref{norma1}) becomes
\be
\Hcal_{ac} \nu^a \nu^c \approx R_{abcd} {\displaystyle{{g^{bd}}\over {\Ncal}}}
{\displaystyle{{g^{ac}}\over {(\Ncal-1)}}} = 
{\displaystyle{{\Re}\over {\Ncal(\Ncal-1)}}}
\ee
where we introduced the scalar curvature of the manifold
\be
\Re \defin R^a{}_a \equiv g^{ac} R_{ac} \equiv g^{ac} g^{bd} R_{abcd} 
\ee
which, using the previous formulae, turns out also to be
\be
\Re \equiv \Sum_{A=0}^{\Ncal-1} \ric({\bf e}_{{}_A}) = 
\Sum_{A=0}^{\Ncal-1} {\Sum_{B=0}^{\Ncal-1}}{}^{'} K^{{}^{(2)}}_{{}_{AB}} 
= \Sum_{A=0}^{\Ncal-1} {\sf Tr}\RcA
\ee
and where we averaged over the right number of directions {\sl pairs}.
Using these approximations, equation (\ref{norma2}) is replaced by
\be
{{d^2z}\over{ds^2}} + k_S~ z = 0
\label{norma3} 
\ee
where the average scalar curvature has been defined, $ k_S\equiv k_S
(\q)\defin {\displaystyle{{\Re}\over {\Ncal(\Ncal-1)}}}$.\\
We see from the derivation itself that the use of scalar related 
quantities is unjustified, because the averaging over directions of 
geodesics means to ignore the actual state of the system; \ie, the
evolution of the perturbation to a {\sl given geodesic, along the geodesic
itself}  results to depend on the paths associated to all the geodesics
passing on the same point \q, with arbitrary velocity! This is the same as 
saying that this claimed {\sl average over states} turns out to be instead 
an average over ${\cal O}(\Ncal^2)$ evolutionary equations amongst which
only ${\cal O}(\Ncal)$ carry meaningful information, which turns out obviously
suppressed by the ${\cal O}(\Ncal^2-\Ncal)$ irrelevant ones.\\ 
We will show below that performing that average amounts to assume that the 
manifold is isotropic; this is definitely not true for whatever realistic 
model of physical \DS, whereas it is exactly what happens for abstract 
geodesic flows of mathematical ergodic theory!\ The only physically meaningful
situation in which the scalar curvature carries the correct information 
is that of two dimensional manifolds, \ie, the case in which there is only
one sectional curvature, equal to the Ricci one, in turn equal to half times
the scalar curvature, \cite{SyngeSchild}.
So, any averaging process must be carried out carefully, if reliable 
results are sought. This is true in general, but in particular, for few 
degrees of freedom systems, there are two opposite situations and a list 
of warnings is mandatory:
\begin{itemize}
\item As remarked above, for two dimensional configuration spaces,
all the curvatures are directly related each other and the results 
obviously do not depend on the choice made.
\item This do not means that for few (but $\Ncal>2$) degrees of freedom, the
scalar curvature should give reliable results. Even more, in these cases,
the global constraints in the geometrical features of the manifold causes
any averaging to destroy the information contained in the sectional curvatures,
causing even the Ricci one to lead to wrong results, as the averaging can 
hide all the mechanisms responsible of the onset of Chaos,
\cite{MTTD,Maratea_HH}.
\item To highlight the peculiarities of two-dimensional manifolds, we remind that
for \DS's with two degrees of freedom, if interested 
only in the {\sl phenomenological} determination of dynamic (in)stability, 
no matter what geometrization is adopted, the results are in agreement. 
But if the goal is the discovery of the origin of Chaos, it should be taken 
into account that a suitable enlargement of the manifold gives deeper 
insights, \cite{CPMT_HH}. This is not necessary if the manifold
is already three-dimensional or more, \cite{MTCP_BIX}.
\item To point out the general (for any $\Ncal$) unreliability of scalar related 
quantities, and even of Ricci curvature in the case of small $\Ncal$, we
recall some results obtained within the Eisenhart geometrization framework,
\cite{Marco93}: for any conservative \DS, $\Re\equiv 0$, $\ric_E(\u)\equiv 2$ 
even for a paradigmatic chaotic hamiltonian (the H\'enon-Heiles system). So,
eq.(\ref{norma3}) turns out to be always meaningless, and even eq.(\ref{norma2})
is completely unreliable for two degrees of freedom Hamiltonians.
\item Also within the Finsler \GDA, it is shown that for a three dimensional 
manifold  (corresponding to a two degrees of freedom \DS, \cite{CPMT_HH}), even 
the use of $\ric(\u)$ leads to incorrect answers. 
\item The predictions on the N-body system
based on the scalar curvature are in complete disagreement with those
obtained using Ricci curvature, which instead agree with those of
standard tools of investigation of chaotic properties.
\item The last points relate again to an hasty extension of results 
borrowed from  Ergodic theory of abstract {\sf DS}'s and to an inadequate 
consideration of {\sl Schur's theorem}. Indeed it has become more and more 
evident that the properties of manifolds associated with realistic physical 
models differ significantly from those of manifolds studied in Ergodic theory 
and that the instability, when present, has very different sources in the two 
classes of geodesic flows. 
\item In the latter class, actually, the mechanism of instability resides in 
the negativity of the sectional curvatures; if {\sl all} the sectional curvatures 
are {\sl always} negative, then the flow is exponentially unstable, the 
system is mixing and all the statistical properties are well
justified, \cite{Katok}. Again, in that case also the Ricci and scalar curvatures 
are {\sl always} negative, and all approximated/averaged equations predict 
instability as the exact ones do.
\item Conversely, the sources of instability for physically meaningful 
geodesic flows are instead related
to the non constant and anisotropic features of the manifold. In such a case 
whatever averaging process risk to destroy both non uniform properties, which 
are moreover each other tightly
related via the Schur's theorem. For \mdf systems simple statistical 
considerations, actually the use of the {\sl Central Limit theorem}, allow 
to recover the fluctuations (both in time and directions) and to obtain 
very accurate answers on the global behaviour, 
\cite{CPTD,MarcoLapo,MarcoLapoLivi}. But, when such statistical arguments 
are not justified by low $\Ncal$ values, the use of averaged quantities 
change completely the answers.
\item Incidentally, we remind that the \GDA~requires some obvious conditions to 
be applied. We refer to \cite{CPTD,MTTD,PRE97a,Aquila2} for an exhaustive 
discussion, and recall here the apparent singularity of the Jacobi line
element, $ds_J$, eq.(\ref{metrica}), at the {\sl turning 
points}, where $\Ucal(\q)=E$. For a {\sl standard} \DS, whose kinetic
energy is a positive definite quadratic form and then these turning points are
only on the boundary of the region allowed to motion, it is well known that 
this do not result in any {\sl real} singular behavior, and, moreover, for 
\mdf systems, the chance that the kinetic energy vanishes becomes smaller 
and smaller with the increase of $\Ncal$. 
Nevertheless, the Jacobi geometrodynamics has been 
applied, \cite{Szydlowsky}, to a three dimensional General Relativistic \DS,
 for which the {\sl kinetic energy} is not positive definite. In that case 
the singularity of the metric is {\sl inside} the region allowed to motion 
 and the Jacobi form of the \GDA~cannot be applied. This is one of the two 
main points which invalidate the results there obtained, in addition to the 
use of the scalar curvature, whose general lack of reliability 
(extolled for two dimensional \DS's) has already been pointed out. 
It is just in order to overcome this kind of limitations
of the Jacobi \GDA, that we proposed the Finsler extension whose application
to Celestial Mechanics and General Relativistic \DS's has been performed 
succesfully, \cite{Sigrav96,MTCP_BIX}. 
\end{itemize}
Summarizing, we can say that, while the use of Ricci related quantities is 
trustworthy for \mdf systems, and for small $\Ncal$ no averaging procedure is
guaranteed to give reliable answers, the scalar curvature has in general,
as it is, no relationship with the issue of instability, except in the very
exceptional situations in which it carries the same informations of the Ricci
or sectional curvatures. 

\section*{Sources of geodesic instability.}
From now on we concentrate on the \GDA~to Chaos in \mdf Hamiltonian systems; 
therefore we will assume that $\Ncal \gg 1$, so we can neglect the full set 
of equations (\ref{EDG}) and limit ourselves to the averaged 
eq.(\ref{norma2}).\\
We are left with a formally simple ordinary differential equation 
for the norm of the perturbation, $z(s)$:
\be
z'' ~ + ~ \chi(s) ~ z ~ = ~ 0\ ,
\label{zsecondo}
\ee
where $(\cdot)'= d(\cdot)/ds$. According 
to the level of approximation, the quantity $\chi(s)$ entering in the evolution 
of $z(s)$ would be, in turn:
\be
\chi(s)~\equiv ~ \tilde\chi~[\q(s),\u(s)] ~ = ~ \left\{ \begin{array}{lll}
k_{{}_R} (s) = \tilde{k_{{}_R}} [\q(s), \u(s)] ~ & = & ~ 
{\displaystyle{{\ric(\u)}\over{\Ncal-1}}} \\ \ & \ & \ \\ 
k_{{}_S}(s) = \tilde{k_{{}_S}} [\q(s)] ~ & = & ~ {\displaystyle{{\Re}\over{\Ncal(\Ncal-1)}}}
\end{array}
\right.
\label{chi}
\ee
From the preceding discussion, it is clear that we will not consider the
second case in eq.(\ref{chi}), except when it is practically equivalent
to the first. Rather, we remind that
another set of ($\Ncal-1)$ equations which, although being approximated, give 
more detailed informations on the stability of the flow, could be obtained 
letting the {\sl frequency} $\chi(s)$ to assume the value of every non trivial 
eigenvalue of $\Hcal^a{}_c$, $\chi(s) = \lambda_{{}_A} (s)$, 
with $A=(1,\ldots,\Ncal-1)$. For systems with few degrees of freedom the 
equations so obtained, yet approximated, have proven to reproduce faithfully 
the qualitative behaviour of solutions of the exact geodesic deviation 
equations, even when the Ricci curvature fails (nothing to say about the 
scalar curvature), \cite{CPMT_HH}.
For high dimensional \DS's the problem stands in the huge computational work 
required to diagonalize $\bfHcal$, rather than in the integration of 
$(\Ncal-1)$ equations instead of one. In some cases, however, the stability 
tensor turns out to 
be easily diagonalizable, \cite{MTCP_BIX}, and the resulting equations 
help to get some hints on the reliability of the results obtained within 
the {\sl Ricci-averaged} approach.\\
Equation (\ref{zsecondo}) is our starting point for the discussion of the 
links between geometry and instability. As it stands, it points out the 
importance of the curvature properties of the manifold on the evolution 
of a generic perturbation to a geodesic flow. 
Later on, we will discuss the steps needed to rephrase the results 
in terms of the
newtonian time $t$; for the time being we limits to general 
considerations on the stability of the geodesic flow, described as the Jacobi 
 $s$-parameter evolves.

Most of the results of abstract Ergodic theory refer to geodesic on manifolds 
of {\it constant negative sectional curvature}, \cite[\S 6.1]{Anosov}. 
In this case, \ie,
when $\lambda_A(s)\equiv -\alpha^2 = {\rm const.}\ ,\  
\forall\, A=1,\ldots,(\Ncal-1)$,
whatever choice is made in eq.(\ref{chi}), it is apparent that it is also
\be
\chi(s) ~ \equiv ~ - \alpha^2 ~ = ~ {\rm const.}
\ee
Well then, eq.({\ref{zsecondo}) implies that 
the magnitude of the deviation will increase exponentially fast in $s$:
\be
z(s) = z_o \cosh (\alpha s) + {{z'_o}\over {\alpha}} \sinh (\alpha s)\ \  
\vainf{|s|} \ \  C~ \exp{\alpha|s|}\ .
\ee
In this case it has been {\sl rigorously proved}, \eg,\cite{Katok}, that the 
dynamics possess the strongest statistical properties, as positive Kolmogorov 
entropy, mixing, exponential decay of correlations, and so on; and it is 
in fact the {\sl most chaotic} one.\ Therefore, if a \DS~were characterized 
by {\sl constant negative sectional curvatures}, then it would be
strongly chaotic and approaching the equilibrium state within a {\sl 
relaxation time}
of order $ S_r\approx\alpha^{-1}$, or, in terms of the newtonian 
time\footnote{A more careful evaluation of the {\sl physical} relaxation 
time would give the result
$${ { \langle\Wcal\rangle_{{}_t} } \over { \langle\Wcal^2\rangle_{{}_t} } }
\minord \alpha\cdot T_r \minord \langle\Wcal\rangle_{{}_t}^{-1}\ .$$ 
Although the two limits do not differ significantly for {\sl standard} 
\mdf systems, as the \FPU chain, in the case of very inhomogeneus 
gravitational N-body system out of {\sl global virial equilibrium}, 
the difference can be quite remarkable. This has no consequences in this 
context because both them have neither constant or negative sectional 
curvatures. The implications of these more accurate limits will be discussed 
elsewhere.},
\be
T_r \approx \left( \langle \Wcal \rangle_t ~\alpha\right)^{-1}\ .
\ee   
As far as we know, in the case of {\sl non constant} but {\sl everywhere 
negative} sectional curvatures, 
\be
\lambda_A(s) \leq - \beta^2 < 0,\ \ \ \forall A=1,\ldots,(\Ncal-1);\ \ \ \ 
\lambda_0 = 0\ ,
\ee
while the flow is definitely unstable, \cite{Katok}, for what concerns the
statistical properties, it is plausible, and there are convincing 
argumentations, according to which also the approach to equilibrium 
take place on a $s$-time scale of order $\beta^{-1}$, although we are not 
aware of any rigourous and explicit theorem.\\
Bearing in mind that the exact equation for the evolution of geodesic 
deviation is eq.(\ref{norma1}), of which eq.(\ref{zsecondo}) is an 
approximation, we should discuss at this point the relevance of Schur's 
theorem (see, \eg, \cite[\S 4.1]{SyngeSchild}), which refers to {\sl 
isotropic} Riemannian manifolds\footnote{The extension to more general 
differential manifolds, as the Finsler ones, is discussed in \cite{Rund}.}, 
whose definition is the following:
\begin{quotation}
"An $\Ncal$-dimensional ($\Ncal>2$) riemannian manifold, ${\sf M}_{\Ncal}$, 
is said isotropic at a point
$\q\in {\sf M}_{\Ncal}$ if the sectional curvatures $K^{(2)} (\x,\y)$ 
evaluated at $\q$, do
not depend on the pair $(\x,\y)$, \ie, on the {\sl 2-plane}."
\end{quotation}
In this case the Riemann tensor has the form
\be
R_{abcd} ~ = ~ {\cal K}\ (g_{ac} g_{bd} - g_{ad} g_{bc})
\ee
where ${\cal K}$ is a scalar quantity, a priori depending on the
position, ${\cal K}={\cal K}(\q)$. 
Well, then the Schur's theorem assures that:
\begin{quotation}
"If an $\Ncal$-dimensional ($\Ncal>2$) riemannian manifold, 
$M_{\Ncal}$, is 
isotropic in a region, then the riemannian curvature is constant throughout 
that region"
\end{quotation}
That is, the scalar ${\cal K}$ is a constant quantity, do not depends on \q. 
This incidentally 
implies also that the covariant derivative of Riemann tensor vanishes.\\
What this theorem entails on the issues we are facing on can be deduced 
looking rather at
its {\sl reverse} implications. Indeed, according to the theorem,
if a manifold is isotropic, then {\sl all} the sectional curvatures are also 
constant; nevertheless, if the manifold is anisotropic, nothing can be said 
about the constancy or variability of the sectional curvatures, as it could 
be either\footnote{In the sequel we will rephrase these argumentations in terms
of the principal sectional curvatures along the flow, (the $\lambda_A$'s), 
though these depend on a parallely propagated vector. This is legitimate as
long as \bfHcal~satisfies very generic hypotheses. A very interesting 
account on these general arguments can be found in \cite{Berndt}.} 
$$
K^{(2)} (\x_1,\y_1) = \alpha_1 = {\rm const.}\ \neq\ K^{(2)} 
(\x_2,\y_2) = \alpha_2 = {\rm const.}\ ,
$$
or $ K^{(2)} (\x,\y) = A(s)$, a variable quantity;
and in this case the theorem do not give more informations. 
Then, this occurrence can 
be interpreted saying either that $\lambda_A\neq\lambda_B$ in consistent 
with both variable or constant $\{\lambda_A\}$, or, conversely, that 
$\lambda_A = {\rm const.},\ \forall\ A$ is consistent with both isotropy 
or anisotropy.\\
Other interesting {\sl back implications} of the theorem explain what can 
be learned looking at the behaviour of Ricci curvature.
\begin{itemize} 
\item If {\sl at least one} of sectional curvatures varies, $\lambda_A=
\lambda_A(s)$, then the manifold cannot be isotropic.
\item For an isotropic manifold, \ie~for $\lambda_A\equiv\alpha,\ 
\forall A=1,\ldots,(\Ncal-1)$, then also the Ricci and scalar curvatures 
are also constant, and $k_R=k_S=\alpha$. 
This is the sole situation in which the averaged quantities equal exactly the 
the sectional curvatures.
\item If the manifold is anisotropic, then the Ricci curvature along 
different geodesics assumes distinct values, whereas the scalar curvature 
do not distinguish between various initial conditions. 
\item Even in an anisotropic manifold, if the sectionals are constant,
then both $k_R$ and $k_S$ do not vary. In this case they represents their 
averaged values.
\item When at least one of the sectionals varies, the manifold must be 
anisotropic, and in this case the averaging procedure can hide the 
fluctuations: with fluctuating $\{\lambda_A(s)\}$,  it is possible to have 
either $k_R(s)$ also varying with $s$, or even
constant. These remarks apply with even more relevance for scalar related 
quantities, \cite{Marco_Nbody}.
\item In particular, if we find that $k_R={\rm const.}$, we cannot say 
nothing about the behaviour of sectional curvatures, because this is 
consistent with all the possible features, namely, we can have either an 
isotropic or an anisotropic manifold with constant sectionals, or even an 
anisotropic one with fluctuating $\lambda_A$'s!
\item Vice versa, if the outcome is that $k_R(s)$ is a fluctuating quantity, 
we can  definitely assert that also the sectional curvatures vary along the 
geodesic, and, by the Schur's theorem, that the manifold is actually 
anisotropic.
\end{itemize}
{\sl Luckily enough}, almost all physically meaningful geodesic flows fall into 
this last category, and either farther from complete integrability we are 
or higher is $\Ncal$, more true it is. According to this picture, the 
manifold is anisotropic, with sectional curvatures rapidly fluctuating, 
almost independently from each other, in such a way that, when $\Ncal\gg 1$,
we can assume (and even verify!) that $\ric (\u)$ is the sum of $\Ncal-1$ 
uncorrelated quantities, of which $k_R (\u)$ represents the average. For 
\mdf systems, and when the motion is definitely far from quasi-periodicity, 
it can be shown,
\cite{CPTD,MarcoLapo,MarcoLapoLivi}, that it is consistent also to assume 
the random character
of the quantities entering $\ric (\u)$, and using the central limit theorem, 
to relate the fluctuations of $k_R(\u)$ to those of the $\lambda_A$'s, 
obtaining an analytical estimate of the asymptotic rate of growth of the 
norm of the deviation vector, \ie, of the {\sf LCN}.\\
Once realized that the case in which {\it all} the sectional curvatures 
are {\sl constant} is an exceptional one, and consequently that 
the rigorous results on abstract geodesic flows carry very little 
physical interest, we turn to the study of the evolution of a 
perturbation to a realistic geodesic flow. 

\subsection*{Mechanisms responsible of the onset of Chaos.}
Bearing in mind the warnings made above, we now look at eqs.(\ref{zsecondo}) 
and (\ref{chi}), in order to see under what conditions instability can arise. 
It is obvious that when $\chi(s) = -\alpha^2 = {\rm const.}$, these equations 
predict an exponential growth of the magnitude of the geodesic deviation. 
As stressed before, a constant value of $\ric(\u)$ can occur for very different 
behaviour of sectional curvatures, but a negative constant value implies 
that the average value of them is permanently negative, and this do not 
happen for any geodesic flow of physical significance, neither it found that 
$k_R$ is mostly negative. To see this, let us write the expressions for 
Ricci and scalar curvature in the Jacobi metric for a conservative \DS~
with $\Ncal$ degrees of freedom, \cite{CPTD}:
\be
k_R(\u) = {\displaystyle{{1}\over{2\Ncal\Wcal^2}}}
\left\{
\Delta \Ucal + {\displaystyle {{(\grad \Ucal)^2}\over {\Wcal}}} 
+(\Ncal-2)\left[ \half \left( {\displaystyle {{d\Ucal}\over{ds}}}\right)^2
+\Wcal {\displaystyle {{d^2\Ucal}\over{ds^2}}} \right]
\right\} \ ,
\label{kR}
\ee
and 
\be
k_S(\u) = {\displaystyle{{1}\over{\Ncal\Wcal^2}}}
\left[
\Delta \Ucal - \left(  {\displaystyle {{\Ncal- 6}\over {4}}}
\right)
{\displaystyle {{(\grad \Ucal)^2}\over {\Wcal}}} 
\right] \ ,
\label{kS}
\ee
where we indicated the usual {\sl $\Ncal$-euclidean} Laplacian and gradient 
operators with
$\Delta$ and $\grad$, respectively.
A first look at the expressions shows that, on passing from the average over 
directions of the deviation vector to the loose approximation given by the 
scalar curvature, some important information get lost. Indeed in both $k_R$ 
and $k_S$ appear the $\Ncal$-dimensional laplacian
of the interaction potential, with a weight reduced by a factor 2 in the former, 
and the 
$\Ncal$-dimensional squared gradient, \ie, the sum of the squared forces 
acting on each {\sl particle} of the system. One main difference is on the 
sign and on the weight of this term, which is amplified by a factor 
${\cal O}(\Ncal)$ in the scalar related frequency, $k_S$ to which it 
contributes negatively, thus increasing the chance of trivial instability 
of the solution. At least equally relevant to the {\sl evolution of the 
instability} 
is the lacking in $k_S$ of the last two terms in $k_R$, as they describe 
the {\sl explicit} $s$-time dependence of the curvature of the manifold, 
in particular for systems out of {\sl global virial equilibrium}, 
\ie, subject to {\sl collective oscillations} which involve macroscopic 
fluctuations of total potential and kinetic energies, as it occur for a 
collisionless N-body system during its very early stages.
The consequences of the differences just listed can be grasped intuitively 
if rephrased in terms of the newtonian $t$-time. As a matter of fact, we see 
that the last two terms in the \rhs~of eq.(\ref{kR}) are nothing else than 
the {\sl logarithmic time derivatives} of the
kinetic (or potential) energy:
\be
{ \displaystyle {{d\Ucal}\over{ds}} } = {\displaystyle { {1}\over{\sqrt{2}} } }
{ \displaystyle {{\dot{\Ucal}}\over{E-\Ucal } }} = - {\displaystyle 
{{1}\over{\sqrt{2}}}}
{\displaystyle {{\dot{\Wcal}}\over{\Wcal}}}\qquad ;\qquad\qquad 
{\displaystyle {{d^2\Ucal}\over{ds^2}}} = - {\displaystyle {{1}\over{2\Wcal}}}
{\displaystyle {{d}\over{dt}}} {\displaystyle 
\left({{\dot{\Wcal}}\over{\Wcal}}\right)} = -
{\displaystyle {{1}\over{2\Wcal}}}
\left[ {\displaystyle {{\ddot{\Wcal}}\over{\Wcal}}} - 
{\displaystyle \left({{\dot{\Wcal}}\over{\Wcal}}\right)^2}\right]\ ;
\ee 
and, for a \mdf system, these relative fluctuations are damped by a statistical
factor that obviously increases with $\Ncal$. Moreover, they exploit an explicit
dependence of $k_R(s)$ on rapidity with which the {\sl global virial 
equilibrium} is attained. The magnitude of those relative fluctuations indeed, 
though generally smaller and smaller as the number of degrees of freedom grows, 
depends in general on the {\sl virial ratio}, and could be not negligible, 
in particular for a self-gravitating system, where collective effects are
likely to occur, due to the long range nature of the interaction. 
Thus, whereas the Ricci curvature keeps memory of these evolutionary processes, 
the scalar one, just because forget {\it ab initio} the real dynamics of the 
flow, erasing the $\u$ dependence, cannot take care
of the consequences of the approach to this dynamical global equilibrium. 
If we add to these
remarks the presence of $(\Ncal-1)$ {\it Ricci curvatures} along directions 
extraneous to the dynamics, the {\sl wrong} sign and weight 
it attributes to the square average force\footnote{Which has been responsible 
of a largely unjustified debate in the past, \cite{GS,Kandrup}.}, with respect 
to Ricci curvature in both Jacobi and Finsler cases, \cite{CPTD,MTTD,PRE97a}, 
together with the evidence of the effectiveness of the use of the latter in 
gaining deep insights on the sources of instability and in the analytical 
computation of instability exponents, \cite{MarcoLapoLivi,CPTD,CPMT_HH}, 
it turns out once more why the use of scalar related quantities is ever more 
unreliable in the case of \mdf systems.\\
A first look to the expression of the average curvature in the $(\Ncal-1)\  
2$-planes
containing $\u$ leads to the following considerations:
\begin{itemize}
\item The $\Ncal$-dimensional laplacian is almost everywhere positive for 
every confining
interaction. It is always positive near the minimum of any potential and 
can be locally negative only in limited regions.
\item The squared $\Ncal$-gradient contributes with a positive sign to $k_R$.
\item For $\Ncal>2$ also the square of the of the $s$-time derivative of the 
potential energy contributes with a plus sign.
\item Then, the term with the second derivative of $\Ucal$ alone can give a 
negative contribution to the Ricci curvature. 
\end{itemize}
What it has been found for the \FPU chain, either  numerically integrating 
the trajectories, \cite{Marco93} and \cite{CPTD}, or canonically averaging 
over phase space, \cite{MarcoLapo} is that the Ricci curvature (and then $k_R$) 
is mostly positive, that its average value is {\it always} positive, that 
there is no correlation at all with the degree of stochasticity in the 
dynamics (as measured by, \eg, the maximal \lcn),\cite{Arg95}, that 
the chances to find a negative value for it decrease quickly increasing $\Ncal$, 
vanishing in the thermodynamic limit. Moreover, in \cite{CPTD}, 
it has been found the numerical evidence of the actual occurrence of 
the foreseen {\sl virial transition} process, \cite{CPPG_nc94,CPPG_Gen}.
In the figures \ref{fig_N-dilogt} and \ref{fig_N-dit}, 
this transition is neatly detected, as well as
its dependence on the number of degrees of freedom and on the
regime of chaoticity. 
It is found that, while during the first phase of {\sl virialization}, 
the frequency of events such that $k_R<0$ is not negligible, after then,
and already for relatively small values of $\Ncal$ ($\sim 10^2$),
the probability $\wp$ of a negative value for $k_R$, is actually vanishing,
\be
\wp (k_R<0) \sim {C_\Ncal}\exp{[-(t/T_v)^b]}
{\raisebox{-1.1ex}{$\stackrel {\displaystyle {\llongrightarrow } }
{\scriptscriptstyle{t > T_v }}$}}\ 0\ ;  
\ee
as highlighted,
from the {\sl pure} $t^{-1}$ behaviour of the 
{\sl measured cumulative frequency}, $F_{-}(t)$\footnote{Defined as the
ratio between the total number of occurrence of negative values of Ricci 
curvature and the total number of steps performed. That is,
$N_{\rm steps}= t/\Delta t,\ \ \Delta t$ being the integration time step.}:
\be
 F_{-}(t) \defin {{N_{-}(t)}\over {N(t)}} = { {N(k_R<0)} \over {N_{\rm steps}} }
\ee
shown in figure \ref{fig_N-dilogt}. 
We observe, moreover the strong decrease of this frequency with $\Ncal$, and
the very weak correlation with the degree of stochasticity, as shown
by figure \ref{fig_N-vsLyap}.
The following discussion explain the 
qualitative behaviours of the constant $C_\Ncal$ and the virialization
time $T_v=T_v(\Ncal,\beta\epsilon)$.
When it was firstly detected, it was surprising enough 
to see the lack of correlation between the degree of stochasticity, as
measured by the maximal \lcn, or the parameter $\beta\epsilon$, and
the frequency of occurrence of negative values. Even the $\Ncal$ dependence
was contrary to what expected on the basis of the belief that the statistical
description is more reliable as the number of particles increases.
On the light of the analytical estimates, since then these results turn 
out however completely understandable. 
A thorough analysis of these results is presented in \cite{CPTD} and
completed in \cite{PRE97b}, but we can say here that the efficiency of
the virialization process is clearly decreased by the occurrence of a
quasi periodic behaviour, \ie, by a nearly integrable dynamics. Moreover,
the {\sl phase mixing}, which is in turn responsible of the trivial loss
of correlations between particles, becomes faster along with the increase
of the largest normal mode frequency excited, \ie, with $\Ncal$ too.\\
As the unpredictability is believed to growth with the numbers 
of interacting particles, the inverse correlations between $\Ncal$ and 
$\wp (k_R<0)$, once more force us to discard
the hypotesis of an instability driven by negative values of Ricci curvature. 
Besides, the results obtained within the Jacobi picture are in complete 
agreement with those coming out from the Eisenhart geometrization,
\cite{MarcoLapo,CPTD}; and in this last framework, the Ricci curvature 
for the \FPU chain is always positive (more, it is always $k_{R_{\rm E}}\geq 2$), 
nevertheless, is there possible to recover in a 
very elegant and effective way all the (in)stability properties
of the dynamics, obtaining an analytical algorithm for computation 
of the largest \lcn.\\
If the sign of $k_R$ is mostly positive, and seeing that numerical 
integrations of eq.(\ref{zsecondo}) allow to recover all the qualitative 
and quantitative stability properties of \mdf systems, we are led to ask 
where stems from the exponential growth of the geodesic deviation. 
The answer is actually written in any textbook, \eg,\cite{LL1}, and resides 
in the mechanism of the swing: the instability is driven by {\sl parametric 
resonance} induced by the fluctuations of the positive value of the 
{\sl frequency} $\chi(s)$. 
We refer to \cite{CPTD,Marco93,MarcoLapo,MTTD,PRE97a} for details, and 
here simply recall that in the non-autonomous equation
\be
\ddot{x} + \Omega^2 (t) x = 0\ ,
\ee 
with 
$$
\Omega^2(t) = \Omega_o^2 [1+ f(t)]\qquad ; \qquad\qquad |f(t)|<1
\qquad ; \qquad\qquad  \langle f(t)\rangle_t = 0
$$
instability (\ie, exponential growth) of the solution can arise, fixed the 
average frequency $\Omega_o$, if suitable conditions are fulfilled by the 
modulation factor $f(t)$.
In the case of eq.(\ref{zsecondo}), the modulating factor is not an explicit 
function of
$s$-time, rather depends on $s$ through the actual {\sl state} $[\q(s),\u(s)]$, 
of the system. 
So, in order to foresee the behaviour of the deviation vector it is important 
to know not only the average value $\chi_o\defin\langle\chi(s)\rangle_s$, but 
also the amplitude, $\sigma^2_{\chi}=\langle (\chi(s)-\chi_o)^2\rangle_s$ and 
the variability time-scale $\tau_s$ of its fluctuations.\\
Once again, we see why the cancellation of the {\sl fluctuating} terms
consequent to the averaging over \u, is at the origin of the unreliable 
results obtained with scalar curvature.

In order to compare the outcomes obtained within the \GDA~with those resulting 
from standard
dynamical system approach, we need to rewrite the equations in terms of the 
newtonian $t$-time,
exploiting the non affine relation given by eq.(\ref{metrica}). 
If we do this,  equation (\ref{zsecondo}) becomes:
\be
{{d^2z}\over{dt^2}} - {{\dot \Wcal}\over {\Wcal}} {{dz}\over{dt}} + 
{\hat{k}_R} z =0\ ,
\label{damped} 
\ee
where $\hat{k}_R\defin 2\,\Wcal^2\,k_R$ is a sort of {\sl rescaled} Ricci 
curvature. We are left wiht an ordinary differential equation for a 
{\sl damped} harmonic oscillator, whose peculiarities lie both in the fast 
variability of the {\sl frequency} $\hat{k}_R(t)$, and in the 
undefined sign of the {\sl damping} term, which actually has zero average 
and can act alternatively also as an {\sl anti-damping}. The presence of this 
term is nevertheless crucial in order to understand
the sources of instability in the {\sl physical time gauge}.
Indeed, an equation completely equivalent to (\ref{damped}) is obtained 
with the change of variable
\be
Y(t)\defin \Wcal^{-1/2}(t)~z(t)\ ,
\ee
where it is understood that all the quantities depends on $t$ through the 
one-to-one correspondence $s=s(t)$, fixed by 
$ds=\sqrt{2}\,\Wcal\,dt,\ \Wcal>0$.\ For \mdf systems indeed the 
probability that $\Wcal=0$ is negligible, 
and, for a confined \DS, the virial theorem guarantees also that $\Wcal(t)$ 
is a quasi periodic function, bounded from above\footnote{This is not 
strictly true asimptotically for the {\it mathematical} gravitational 
N-body system, but it is true for almost all initial conditions at any 
finite time. Moreover, and this is an important point, $\Wcal$ is 
{\sl always bounded} from above, when the {\it physical} 
interaction is addressed. We stress that the strategy we introduce below
to cope with the mathematical singularities of the gravitational potential
{\sl do not change} its very nature of {\sl non-compact} DS.}; 
in such a way that the qualitative long term time behaviours of 
$Y(t)$ and $z(t)$ are the same.
This well known substitution allows to get rid of the damping term, 
leading to a standard
{\it Hill's equation}:
\be
{\ddot Y}(t)~ + ~ Q(t) ~ Y(t)~ =~ 0\ , 
\label{pupat}
\ee
where the frequency $ Q(t) $ follows from the rescaled average curvature 
$ {\hat{k}_R}(t)$:
\be
Q(t)\equiv {\tilde Q}\{\q (t)\} \defin {\hat{k}_R}(t) - {3\over 4} \left({{\dot
\Wcal}\over {\Wcal}} \right)^2 +\ 
{1\over {2}} {{\ddot \Wcal}\over {\Wcal}}.
\ee
As heralded above, the terms appearing in the expression of $ Q(t) $ with 
respect to  
$ {\hat{k}_R}(t) $ are very important for the understanding of the behaviour 
of the perturbations in the {\sl physical} time.
To see why, let us write down the explicit $t$-expressions:
\be
{ \hat{k}_R}(t) ={{1}\over{\Ncal}} \left\{ \Delta\Ucal +
{{(\grad {\Ucal})^2}\over {\Wcal}} +
\left( \Ncal -2\right) \left[ {3\over {4}} \left( {{\dot\Wcal}\over
{\Wcal}} \right)^2\ - {1\over {2}} {{\ddot\Wcal}\over {\Wcal}} \right]\right\}\ , 
\label{kRt}
\ee
\be
Q(t) = 
 {{1}\over{\Ncal}} \left\{ \Delta\Ucal +
{{(\grad {\Ucal})^2}\over {\Wcal}} +
\left[ {{\ddot\Wcal}\over {\Wcal}} - {3\over {2}} \left( {{\dot\Wcal}\over
{\Wcal}} \right)^2 \right]\right\}\ .
 \label{Qdit}
\ee
We see that the terms arising from the $s$ derivatives of the potential energy, 
and here
rewritten in terms of $t$ derivatives, appear in $Q(t)$ with a weight that, in 
the large $\Ncal$ limit, is actually reduced by a factor $\Ncal^{-1}$. 
The relevance of this outcome becomes self-evident as we consider the the 
sources of instability in eq.(\ref{pupat}), or (\ref{damped}). Indeed, as 
long as we consider {\sl customary} \mdf systems, as emphasized above, it 
is {\it never} found that $Q(t)<0$, and the frequency with which $\hat{k}_R(t)<0$, 
is always very small, becoming absolutely negligible with increasing $\Ncal$, 
in which case asimptotically {\sl vanishes}, as the virial equilibrium is 
attained; see figures \ref{fig_N-dilogt} and \ref{fig_Qdit200} and 
also \cite[\S 5.2.1]{CPTD}.
In addition, {\sl this frequency do not show any correlation 
with the degree of stochasticity}, as it is evident from figure 
\ref{fig_N-vsLyap}, obtained from \cite{CPTD}, see also \cite{Arg95}.\\
The occurrence of Chaos, \ie, the exponential growth of all the solutions 
(except, possibly, for a set of initial conditions of zero measure) of 
eq.(\ref{pupat}) is consistent with a positive sign-definite $Q(t)$ if 
the mechanism responsible of the onset of instability is the {\sl parametric 
resonance} mentioned above. It is crucial for this phenomenon  to occur 
that the amplitude of fluctuations fulfil some conditions,
whose details depend either on their quasi periodicity or on their
stochastic behaviour. This explain the relevance of the reduced weights in 
front to the fluctuating terms in $Q(t)$ with respect to $\hat{k}_R(t)$ in order 
to recover the known results in terms of the newtonian time.\\
The argument just discussed enlighten also why a \mdf system should go 
across a {\sl transition} in its qualitative behaviour along the approach 
to the {\sl global virial equilibrium}, during which the amplitude of the 
fluctuations in its macroscopic parameters, as
the kinetic and potential energies is damped just in consequence of the 
trivial {\sl phase mixing} previously recalled\footnote{
By this expression we mean the purely kinematic effect of 
particle-particle correlation loss, entailed by the different 
{\sl orbital times}, not to be confused with the {\sl mixing} property of 
Ergodic theory. 
This explain also because very near to integrability, being the motion 
quasi periodic, such correlations persist in time, decreasing even the speed 
of this {\sl phase mixing}, which is in turn responsible of the convergence 
of the curvatures to their asymptotic values. 
We already discussed  
the increasing effectiveness as the {\sl thermodynamic limit} is approached. 
These remarks on the relevance of $\Ncal$ are intended to warn about the
risks hidden in extrapolating naively the outcomes of numerical simulations
performed with too small $\Ncal$; in particular for non extensive interaction
potentials, like the gravity, where, as we will show, the $\Ncal$-dependence
is often hardly predictable.
In these situations, the numerical evaluation of average values and
dispersion of geometrodynamical quantities should be performed with long 
enough runs involving a suitably extended $\Ncal$ range, to single out
possible scaling behaviour.\label{phasemix}}.
When the virialization phase 
is accomplished, the instability of the dynamics is governed mostly by the 
fluctuations in the {\sl dominant} term, which, accordingly to a kind of
{\sl principle of equivalence of any geometrization} (in any case 
confirmed by analytical calculations and numerical computations) is the 
$\Ncal$-dimensional euclidean laplacian, see table \ref{tab_stime}. 

Being the translation of the dynamics in (Jacobi) geometrical language almost 
straightforward for any additive interaction potential, $\Ucal$, we could 
achieve order of 
magnitude estimates of all these quantities in the general case; nevertheless, 
to avoid a (still) heavier discussion, we will consider separately the \FPU 
unidimensional chain and the gravitational 
N-body problem\footnote{We will restrict to the physical case of three spatial 
dimensions, although many theoretical hints on the issue of relaxation in 
N-body systems have been achieved studying two or (mostly) one-dimensional 
systems, which display the very nice property to allow for computer simulations 
virtually free from numerical errors, \cite{Tsuchiya}.}, that can be considered 
most representative among the short and long range, respectively, interaction 
potentials. 

\subsection*{{\sl Standard} potentials: the {\sf FPU} chain.}
The first \mdf system investigated by computer simulations has been the 
celebrated unidimensional chain of N weakly anharmonic oscillators. 
According to the power-law of the anharmonic terms, cubic or quartic, 
these models were indicated as {\sf FPU}-$\alpha$ and
{\sf FPU}-$\beta$ models, respectively, \cite{Fermi}. We will focus our 
attention here on 
the second kind, essentially because it has been thoroughly investigated, 
both from dynamical as well Statistical Mechanical points of view, 
see, \eg, \cite{Marco91,GalganiFPU} and references therein.\\
In this case $d=1$, so $\Ncal=N$ and the Hamiltonian is 
\be
H(\x ,\p ) =\sum_{i=1}^N\left[ {1\over 2}p_i^2+{1\over 2}(x_{i+1}-x_i)^2+
{\beta\over 4}(x_{i+1}-x_i)^4\right]
\label{FPU_Ham}
\ee
We recall that, in the quasi-harmonic limit, namely for $\beta\epsilon\ll 1$, 
where
$\epsilon\defin E/N$ is the specific energy ($E\equiv H(\x,\p)$), the 
introduction of the {\sl normal modes} allows to exploit the properties 
of this quasi-integrable system. Such a {\sl good} set of coordinates, 
$({\bf X},{\bf P})$, is defined as, \cite{TKS}: 
\be 
x_i=\left( {2\over N}\right)^{1/2}\ \Sum_{k=1}^{N-1}
X_k\sin\left( {{i k\pi}\over{N}}\right)\qquad\ ,\qquad\qquad \  P_k = \dot {X}_k 
\ee 
and is well suitable for a {\sl perturbative} treatment. 
The Hamiltonian, eq.(\ref{FPU_Ham}), in the new coordinates reads
\be
H({\bf X},{\bf P})=\sum_{k=1}^{N}\left[{1\over 2}({P_k}^2+{\omega_k}^2 {X_k}
^2)\ \ +\ \beta\!\Sum_{\{ {j_1}, {j_2}, {j_3}\} =1}^N C(k, j_1, j_2, j_3)X_k
X_{j_1}X_{j_2} X_{j_3}\right]  
\ee
with  $ \omega_k = 2\sin \left({k\pi}\over N\right) $  and the
$ C(k, j_1, j_2, j_3) $ are coefficients depending on the choice of the boundary
conditions (fixed or periodic). Using these coordinates the harmonic limit,
$\beta\epsilon\rightarrow 0$, 
apparently turns out to be integrable (actually $N$ independent harmonic 
oscillators!), and the presence of the anharmonic term is responsible for 
the coupling of the normal modes, and the loss of integrability.\\
However, since we are interested in the behaviour when the system is very 
far from integrability, the use of the normal modes coordinates, although 
enlightening for the analytical estimates at low $\epsilon$, is of no help 
in the {\sl strong stochasticity} regime, and it is superfluous from a 
numerical viewpoint, as it would imply a reduced efficiency of the 
algorithm.\\
To estimate the order of magnitude of various terms entering the Ricci curvature, 
we start inquiring on the {\sl dynamical time scale} of the system. This is 
relevant to our analysis
as it enters in the $s$ derivatives  of the potential or kinetic energies.\ 
As most of interaction potentials, the \fpub model possess in the {\it genoma} 
its own lenght and time scales. These generally depend on some global 
parameters describing the macroscopic state of the system 
(as temperature and density). This \DS~is however peculiar in this respect 
being {\sl isochronous} at low energy, as any harmonic system; 
\ie, when $\beta\epsilon\ll 1$, the global dynamical time-scale is of order 
unity.\ An important and well known property of 
this model, is the existence of a {\sl strong stochasticity threshold} 
({\sf SST}) as the specific energy $\epsilon$ increases above a
{\sl critical value} $\epsilon_c$, being $\beta$ held fixed, distinguishing 
among two different {\sl regimes} of Chaos, and also involving a
transition on the features of the relaxation processes driving the system 
towards the equilibrium state, \cite{Marco91,Marco93}.
This \SST~is easily located by the intersection of two different scaling 
law of the maximal \lcn, $\gamma_1$, with $\epsilon$: the first results, 
reported in the references above, seemed to support the claim that
\be
\gamma_1 (\epsilon) \propto \epsilon^2,\ \ \ {\rm for}\ \ \ 
\epsilon\minord\epsilon_c\qquad ;\qquad\qquad
\gamma_1 (\epsilon) \propto \epsilon^{2/3},\ \ \ {\rm for}\ \ \ 
\epsilon\maggord\epsilon_c\ ;
\ee
where $\epsilon_c\simeq\beta^{-1}$. A series of numerical computations with a 
large enough $N$ and carried out for very long integration intervals, 
\cite{CPTD}, extended the  previously investigated range of specific energies, 
and while confirming the power law at low energies, find out that above the 
\SST, the scaling is
\be
\gamma_1 (\epsilon) \propto \epsilon^{1/4},\ \ \ {\rm for}\ \ \ 
\epsilon\maggord\epsilon_c\ .
\ee
These outcomes are in complete agreement with the elegant \SM canonical 
calculations, performed in the thermodynamic limit, in 
\cite{MarcoLapo,MarcoLapoLivi}.\\
Within this context the \GDA~revealed is effectiveness in pointing out 
this threshold in a straightforward manner: figure \ref{fig_ricdiE} 
shows the behaviour or $\langle k_R(\epsilon)\rangle_s$ 
for different values of anharmonicity. The transition energy is neatly 
detected and it should be remarked the definitely quick convergence of 
the Ricci curvature to the average values there reported, for all values 
of $\beta\epsilon$, as it is shown in \cite[\S 5.3]{CPTD}.\\
Moreover, within the {\sf GDA}, it has become feasible to use a semi-analytic 
computation of the largest \lcn, performed in \cite{CPTD} using the computed 
time average, dispersion and correlation time of the Ricci curvature, and in 
\cite{MarcoLapo}, using the same quantities, evaluated instead from a \SM 
point of view, averaging over phase space. 

The corrected scaling law behaviour found above the \SST, though perhaps 
not crucial for what
concerns the {\sl existence} of two different regimes of Chaos, is relevant 
to our present discussion in that a simple calculation, \cite{CPTD}, shows 
that in the strongly anharmonic regime, the {\sl dynamical time scale}, 
$t_D$, of the \fpub model scales exactly as $\epsilon^{-1/4}$, in such a way 
that the quantity $L_1\defin\gamma_1\cdot t_D$ is constant when strong Chaos 
has developed. This point is of outmost importance for the understanding of 
the nature of instability present in the gravitational N-body system.\\ 
In \cite{CPTD}, where it has been explored the dynamical behaviour 
of the \fpub system also for a range of values of the anarmonicity parameter, 
$\beta$, it has been shown that the relevant quantity that determines the level of 
stochasticity of this \DS~is, as expected, the product $\beta\epsilon$. 
Therefore, in the sequel, we will assume that the anharmonicity parameter 
is fixed (namely, $\beta=0.1$) and let $\epsilon$ to vary. Moreover, all the 
quoted numerical results refer to simulations performed adopting periodic 
boundary conditions, \ie, $x_0 = x_{\Ncal},\  x_{\Ncal+1} = x_1$.

Once established the scaling behaviour of the dynamical time with energy, 
we turn to the order of magnitude estimate of the quantities entering the 
Ricci curvature for the quartic \FPU chain.\\
For example, the explicit computation of the $\Ncal$-laplacian is 
straightforward and yields\footnote{Incidentally, this quantity also
coincides with the Ricci curvature 
per degree of freedom in the Eisenhart metric, cfr. \cite{MarcoLapo}.}
\be
{{\Delta\Ucal}\over{\Ncal}} = 2  + {{6\beta}\over{\Ncal}} \Sum_{i=1}^{\Ncal} 
(x_{i+1}-x_i)^2
= 2 + 6 \beta \langle (\delta x)^2\rangle_{{}_\Ncal}\ ,
\ee
which can be also written as (for the details, see \cite{CPTD,PRE97b})
\be
{{\Delta\Ucal}\over{\Ncal}} \cong 2 \left[ 1 +3\left( \sqrt{1+ \beta\epsilon}
-1\right) \right]\ ;
\ee
whereas for what concerns the dynamical time scale it is easy to find that
\be
t_D = {\displaystyle \left[ { {\sqrt{1+4\beta\epsilon}-1}\over 
{2\beta\epsilon}} \right]^{1/2} } {\cal I}(c)
\ee
where ${\cal I}(c)$ is almost constant dimensionless integral, and $c$ is a 
parameter measuring the departure from harmonic behaviour, and so is
directly related to $\beta\epsilon$: $c=1$ in the harmonic case, $c=0$ in
the purely {\sl quartic} domain. It has to be remarked that ${\cal I}(c)$
is a monotonically increasing function of its argument, whose extremal
values are 
$$
{\cal I}(0) = {{\sqrt{2}\,\pi^{3/2}}\over 
{4\left[\Gamma\left({{3}\over {4}}\right) \right]^2}}\ \cong 1.311\ldots
\qquad\ ; \qquad
{\cal I}(1) = {{\pi}\over{2}} \cong 1.5708\ldots
$$
Similar expression can be found about the $(\Ncal,\epsilon)$
dependence of all other quantities,  as resumed in table \ref{tab_stime}.
In \cite{PRE97b}, where a 
comparative analysis between \fpub model and self gravitating N-body system is
also presented, we propose an 
explanation of the scaling law behaviours of Lyapunov exponents on the basis 
of these estimates, supplemented with analogous evaluations on the amplitude of
fluctuations,  {\sl during} and {\sl after} the {\sl virialization} process.\\
A preliminary remark is devoted to the relation of quantities entering the 
frequency $Q(t)$ (or $\hat{k}_r(t)$) to the dynamical time $t_D$. While at 
intermediate $\epsilon$ values the relationship can't be singled out by eyes, 
in both the asymptotic regimes it is easy to see,
that all the quantities behave as $t_D^{-2}$, the different weight they have 
into $Q(t)$ being solely due to their $\Ncal$ dependences.  
We see from table \ref{tab_stime} that among the {\sl static} 
terms\footnote{This is only a rather generic terminology, to let it be 
understood that these terms do not involve explicit time derivatives.
Of course, they also slowly evolve along with the process of approaching
the global dynamical equilibrium, though in a very different fashion in the two
\DS's we are considering. Nevertheless, the {\sl virial transition} is much 
less evident for them than for the other two terms, in such a way that,
in the \fpub chain as well as in the gravitational N-body system too 
(although with a less clear hierarchy), after the first transients, 
the fluctuations responsible for the onset of Chaos come essentially 
from them alone.\label{perdire}}, 
the one related to the average squared force 
is always depressed by a factor
$\Ncal^{-a}$ with respect to the laplacian, with at least $a\maggord 1$. 
The use of the {\sl central limit theorem} allows then to state that, for 
the \fpub chain, the {\sl fluctuating} terms (see note \ref{perdire}), 
although comparable with the gradient term when the system is out of virial 
equilibrium, are then always negligible, in the large $\Ncal$, with respect to 
$\Delta\Ucal/\Ncal$, and are even damped by a further factor of at least 
${\cal O}(\Ncal^{-1/2})$ or ${\cal O}(\Ncal^{-1})$ when that equilibrium is 
attained. As the term for which the damping is less effective, \ie, that
with the second order derivative with respect to time, has also a
{\sl zero time-average}, we understand why this it is important to fully
appreciate the relevance of this transition in \mdf systems.

On the grounds of the above discussion, we understand why the results obtained 
within the frameworks of the Jacobi and Eisenhart metrics agree for the 
\fpub model: the geodesic instability is driven by the fluctuations of the 
{\sf GDI}'s rather than by their negative values. For the \FPU chain, 
all the terms which appear in $Q(t)$ in addition to $\Delta\Ucal/\Ncal$, 
which is nothing but the Ricci curvature per degree of freedom in the 
Eisenhart geometrodynamics, after the virialization phase has been completed 
can be neglected not only with respect to the average value of the latter, 
but even with respect to its own fluctuations, as the geodesics explore the 
manifold. This is generally true at all regimes, \cite{PRE97b}, and is always 
true above the \SST, where the analytic computations of the maximal
\LCN obtained using some elementary results of the theory of Stochastic 
differential equations, \cite{Vankampen}, coincide, irrespective of the 
metric used. For small values of either $\beta\epsilon$ or $\Ncal$,
the quasi-periodic nature of the motions as well as the reduced effectiveness
of the damping cause a much slower convergence to asymptotic values of
the fluctuations, so reducing the reliability of finite time estimates.

\subsection*{Long range, unscreened interactions: the gravitational 
N-body system.}
When we focus our attention towards the gravitational N-body system, 
we are at once faced with an ambiguity which cannot be solved if considered 
only from a mathematical point of view. Along with physically relevant
{\sl real} peculiarity indeed, the Newtonian Gravity possess some mathematical
singularities which do not actually involve any physical singular
behaviour. So, in translating the dynamics into a geometric picture,
we will try to cope with these mathematical divergencies without modifying
the physically peculiar features of this interaction. 

To describe the dynamics of $N$ bodies of mass $\{m_i\}$ interacting 
via the gravitational potential 
$$ 
V (r_{ij}) = -Gm_i m_j / r_{ij}\ ,
$$
we introduce the coordinates $ \{q^a\} $ in the 3N-dimensional 
manifold $M_{\Ncal} $  
$$
{\cal M} : \{ q^a \} \ t.c.\ {\cal U} (\q) \le  E  ( < 0 )\ ;
$$ 
where:  
\be 
q^{a} = {\sqrt{m_i}} r^{\alpha}_i \qquad ;\qquad   
a =3(i-1)+\alpha \ \
{\rm and\ }\ 
  i= 1,\ldots , N ;\ \  \alpha=1,2,3\ \Longrightarrow\ a = 1,\ldots ,3N =
{\cal N}. 
\ee
endowed with the Jacobi metric, eq.(\ref{metrica}), with the total potential 
energy given by
\be
{\Ucal}(\q) =  \Sum_{i=1}^{N-1} \sum_{k=i+1}^N V (r_{ik})
\equiv  {\displaystyle{1\over 2}} \sum_{i=1}^{N} \sum_{k\neq i} V (r_{ik})
\ee
In order to get rid of unnecessary complications, we assume that all the 
particles have unit mass\footnote{The problems we are now discussing do not 
depend on this point. It is obvious that this unrealistic assumption should 
be removed when issues as the problem of equipartition of energy, and the 
consequent {\sl mass segregation} is addressed. 
But the conceptual question we are focusing on now would only be made 
heavier if a non uniform spectrum of masses is allowed.}, 
in such a way that the total mass of the system is $N$, its average 
density $\bar{\rho}=N/V$, where $V$ is the volume occupied by the points. 
Amongst different possibilities, we selected this choice of units because
it seems to be one of the most reasonable, and moreover allows for a direct 
interpretation of the results recently obtained, \cite{Marco_Nbody}.\\
In analogy with what has been done for the \fpub model, we start to inquire 
about the possible existence of any scaling law of the dynamical time, 
$t_D$, for the N-body problem. This question is a very crucial one, because 
of the well known {\sl scale-free} nature of Newtonian Gravity. Nevertheless, 
an N-body system {\sf do} possess indeed a {\sl natural time-scale}, which is 
known in stellar dynamics as the {\sl orbital time}, because it represents a 
suitable average of the typical period of a body moving in the system.
Neglecting 
some numerical factors of order unity, completely irrelevant for the present 
discussion, this dynamical time-scale is usually quoted in terms of the average 
mass density:
\be
t_D \defin {{A}\over \sqrt{{\sf G}\bar{\rho}}} \propto D^{3/2}/N^{1/2}
\ee
where $D\sim V^{1/3}$ represent the typical spatial extension of the system, 
and we have exploited our choice on the mass spectrum, neglecting again constant 
factors. It is easy to show that $t_D$ represents the right order of magnitude 
for most orbital periods even in a moderately inhomogeneus system, except for 
the small fraction of tightly bound {\sl core objects}.\\
In a recent paper, \cite{Marco_Nbody}, where for the first time were performed 
numerical simulations specifically devoted to support (or even correct) the 
\GDA~to Chaos for 
self-gravitating N-body systems, it has been reported, along with other 
interesting
analyses, the absence of any threshold in the behaviour of the maximal 
Lyapunov exponent along with the increase of the energy density 
$\epsilon\defin\! E/N$. 
Moreover, it has been there discussed
the seemingly non physical behaviour of this system when the number of degrees 
of freedom is varied.
It has been found, indeed, that increasing the number of particles,
 the degree of stochasticity, as measured by the largest \lcn\footnote{For 
a non compact phase space, as that of the newtonian N point masses, it is not 
rigorous to speak about {\it true} LCN's, but we can equally adopt this 
terminology, referring to the paper, \cite{Marco_Nbody} for a critical 
discussion of this item. Besides, we defer to the following discussion 
and to a forthcoming paper, \cite{PRE97b}, where some of the open questions 
will be addressed in a hopefully self consistent way.}, decreases, contrary 
to physical intuition, Statistical Mechanics expectations, and to what
observed in any other \mdf system considered before.
The authors refused to believe in this apparent discrepancy, and attributed 
the cause to the too strong approximations which, starting from {\sl exact} 
\jlc equations of geodesic spread, led to an equation like our 
eq.(\ref{pupat}).
We now show that their numerical results are indeed correct, and that no 
failure can be attributed to the averaging process that lead firstly to 
equation (\ref{norma1})  and down there to eqs.(\ref{zsecondo}) and 
(\ref{pupat}). The seemingly unphysical result, turns out to be instead 
completely reasonable, if we consider how the numerical simulations were 
performed\footnote{Actually these scaling laws, reported in table \ref{tab_stime},
do not depend qualitatively on this particular choice.}. 
As remarked above, all the bodies have been assigned unit mass, 
and initial conditions are generated from a spatially uniform distribution 
whose lenght scale $D$ is  chosen to give the desired value of the total 
energy $E$, being the velocity (components) of the masses 
populated according to a gaussian distribution (in the {\sf CM} coordinates 
system) whose variance, $\sigma_v^2$, is adjusted to fulfil the 
{\sl virial theorem}, \ie,\  $ 2~\langle{\cal T}\rangle\,=\,
-\langle\Ucal\rangle $, so that $D$ and $N$ completely determine the value
of total energy, $E=\langle\Ucal\rangle/2\vainf{N} \Ucal/2$.\\
With this choice of initial configurations, in the case
$N\gg 1$, the total energy ($G=m_i=1$),
\be
E \approx {{\Ucal}\over {2}} = 
-{{1}\over{2}} \Sum_{i=1}^{N-1} \Sum_{k=i+1}^N {{1}\over{r_{ik}}}\ ,
\ee
scales obviously as
\be
E = -{{N(N-1)}\over{4}} \langle r_{ik}^{-1}\rangle \cong - N(N-1) 
{{F_\rho}\over {D}}
\ee
where $F_\rho$ is a factor of order unity depending on the actual density 
distribution, and it is so allowed to evolve slowly with time, along with 
the development of a typical 
{\sl core-halo} structure, although this happens on time scales longer than 
those lasted usually by numerical simulations. So the total energy scales 
essentially as $|E|\approx N^2/D$. What this implies on the relationship 
between specific energy $\epsilon$ and the dynamical time scale $t_D$? 
The answer is straightforward: we have $| \epsilon | \approx N/D$; on the 
other end we have also
\be
t_D = {{D^{3/2}}\over {N^{1/2}}} \approx {{(N^2/|E|)^{3/2}}\over {N^{1/2}}}
 = {{N}\over {| \epsilon |^{3/2}}}\ ;
\ee
which is a result valid at any energy. 
Incidentally, we note that actually $t_D\sim D/\sigma_v$.
This simple calculation explain all the results found in 
\cite{Marco_Nbody}; indeed the scaling 
behaviour of the largest \LCN as $|\epsilon |^{3/2}$, with $N$ 
{\sl held fixed}, is consistent again, as for the \FPU chain 
{\it above the} {\sf SST}, with a constant 
value of the quantity $L_1 = \gamma\cdot t_D$. Even the (un)believed 
{\sl unphysical} 
result of a decreased stochasticity when $N$ increases, turns out
to be consequent to the neglecting of the $N$ dependence of $t_D$, and
its interpretation becomes instead very clear 
within this picture: again, the dynamical time is proportional to $N$, if 
$\epsilon$ is held constant, in such a way that the instability exponent 
decreases in absolute value, but remains constant if measured in units of 
the dynamical frequency. In a sense, the lack of {\it stability} property, 
\cite{Ruelle}, by the Newtonian potential, \ie, the {\sl non-extensivity} 
of the potential energy, reflects itself on a dependence on $N$ of the 
dynamical time scale which do not occur in any customary interaction.
The preceding discussion  so contributes to give to the results presented 
in \cite{Marco_Nbody}
a physical interpretation which largely confirms what argumented there and 
moreover support
in a stronger way the use of the {\sf GDA}, which turns out to be very 
fruitful 
even in the application to this peculiar and singular \mdf system. 
It has been removed in fact the doubt about the  reliability of the 
averaged equation (\ref{norma1}), which really carries out almost all the 
informations about the qualitative behaviour of the dynamics of \mdf 
Hamiltonian systems. The scaling law behaviours found numerically agree 
with a surprising precision with the analysis here proposed, for the LCN 
computation, and also for the evaluation of some quantities, named 
{\sl Geometric Chaoticity Indicators} ({\sl GCI}) by the authors: 
we suggest to look at figures 4 and 6 of the cited paper, to see how both 
the $\epsilon$ and $N$ dependences are perfectly fitted by the analytical 
estimates here presented (we remark that the scales used in the figures
there are in natural logarithms).

Nevertheless, when we try to apply the \GDA~to gravitational N-body system, 
another
issue should be necessarily addressed: whatever geometrization procedure 
is adopted, the term
in the Ricci curvature (or even in the scalar one) which is usually {\sl 
dominating} over the others has in this case a very singular behaviour. 
The laplacian of the Newtonian potential indeed vanishes everywhere except 
on the positions occupied by the sources of the field, where
it diverges. As one of the most successful applications of the \GDA~for 
high dimensional \DS's resides in the possibility
of computation of {\sl time and \SM canonical averages} of geometrical 
quantities entering the evolution of perturbations, used to determine the 
stability of trajectories avoiding the explicit integration of the 
variational equations, it emerges clearly why we need a {\sl recipe} 
to cope with such singularities. This problem is also related to the 
Statistical Mechanical properties of gravitational N-body systems that 
we will not touch here, \cite{PRE97b}.
If we perform a {\sl static} average of the Ricci curvature over {\sf M}, 
we neither need nor can to avoid these
${\cal O}(\Ncal^2)$ singularities, which are nonetheless 
{\sl Lebesgue-integrable}, and do not create any difficulty in the 
definition of the average values $\langle k_R\rangle_M$ or any other related 
quantity; although they contribute heavily to the magnitude of fluctuations 
around these averages. Vice versa, when we look at time averages, we face up 
to a problem of probability: what is the chance that a {\sl true} collision 
among particles do occur? If we have to do with the purely mathematical 
problem, we are without hope: the \GDI~ possess their well defined static 
averages, to which the corresponding time averages never approach! 
We feel to be faced again, from a novel perspective, to the classical 
problem of the non existence of an equilibrium state for the {\sl mathematical} 
Newtonian gravity, dressed within this framework as the {\sl non existence} of
an {\sl ergodicity time}. 

The way out to this problem is, as expected, a softening in the potential. 
This is an usual
{\sl trick} (either implicit or explicit) of N-body numerical simulators, 
adopted to cope with the so-called {\sl close encounters} which occur 
sometimes during simulations. It is traditionally passed on that the 
occurrence of such kind of events is a consequence of the necessarily small 
number of particles used in a feasible simulation. This is true to a limited 
extent, from a numerical point of view, but it is a general occurrence, from 
a physical perspective, that during an orbital period some {\sl few} stars 
suffer of such an event.
We avoid to discuss here the details of the criteria adopted to define such 
a close encounter, \cite{PRE97b}, and limit ourselves to say that when 
the distance between any pair of {\it stars} becomes exceedingly small 
with respect to the average 
interparticle separation, $r_{ik}\ll d = D/N^{1/3}$, the newtonian interaction 
should break down, as other effects, possibly even non conservative, 
take place.\\
The easiest way to take into account for this breakdown consist to modify 
slightly (very very slightly) the interaction potential, allowing for a 
softening parameter, $\varpi$ (we don't use the {\sl usual} $\varepsilon$  
to avoid confusion with the specific energy $\epsilon$) enter in the 
{\sl distance} between particles used to compute the potential:
\be
r_{ik} \defin \left[ (x_k-x_i)^2 + (y_k-y_i)^2 + (z_k-z_i)^2 + \varpi^2 
\right]^{1/2}
\ee
The introduction of the softening, helps to improve the {\sl conceptual} 
understanding of the peculiarities of Gravity without modifying appreciably 
the qualitative overall dynamics, which is determined by the long range 
nature of the interaction, rather than by the short range divergencies, 
involving only a negligible fraction of masses (when $N\gg 1$). 
The non-compact nature of the phase space, the {\sl bad} statistical
properties of gravitational interaction (that of being not a 
{\sl stable} one) are indeed preserved by the introduction of the
softening. This because it do not changes at all the long range behaviour
of the interaction, neither introduces an {\sl hard core}, as it could
be erroneusly understood. 
The presence of the term $\varpi^2$ in the potential has so theoretical 
consequences by far more relevant than the practical ones obtained in 
the more well behaved numerical integration of equations of motion. 
Indeed, this {\sl trick} allows both to evaluate analitically and to 
numerically compute, just for a check on those estimates, everywhere 
non-divergent quantities entering the
\GDI, to find their scaling law behaviour with relevant global parameter 
of the system (\ie, the number of degrees of freedom $\Ncal=3N$ and the 
{\sl energy density} $\epsilon$), by means of which to locate the possible 
relationships with the {dynamical} time-scale, $t_D$.\\
As we will see, the analytical estimates suffice to single out what 
is it the {\sl dominant term} in the {\sl frequency}  $Q(t)$ appearing 
in equation (\ref{pupat}), and also a precise determination of its 
relationship with the energy density $\epsilon$. For {\sl standard} 
potentials, as the \fpub one, we have seen above that, provided 
$\Ncal\gg 1$, neither the dynamical time $t_D$, nor the average values of 
$\hat{k}_R$ and $Q$ depends on $\Ncal$, which weakly affects only 
the amplitude of the fluctuations. As expected, instead, the 
Gravity, ever corrected from too mathematical schematizations, shows a 
very different behaviour, disclosing a very intricate dependence of the 
\GDI~ on the number of particles $N$. This relationship can only be guessed 
through order of magnitude estimates, and we need to appeal to numerical 
simulations, whose results let us to state that:
\begin{itemize}
\item The frequency $Q(t)$ is overwhelmingly positive for any value of 
$N$, so the Chaos observed (or better, the exponential instability) is 
certainly due to parametric resonance.
\item The Ricci curvature $k_R(s)$, and consequently $\hat{k}_R(t)$, is 
sign fluctuating for very small values of $N$, but shows a clear tendency 
to become more and more positive as the number of point masses increases.
Already for some tens of bodies, it turns out to be mostly positive.
\item In any case, the evidence of the scaling law with $N$ of all the terms 
appearing in $k_R(s)$, allows us to claim that, even for relatively small $N$,  
the average value of Ricci curvature is always positive, because the only
term contributing with a constant negative sign is the faster to decrease 
with $N$.
\item The evidence of a {\sl virial transition} with the implied damping
of fluctuations.
\item If, moreover, the question of equivalence among static and dynamic 
averages is addressed, taking so into consideration the laplacian term, 
which is always the {\sl biggest} one in the average, irrespective 
on the detailed way it is accounted for, the scaling law behaviour with $N$, 
prove that the Ricci curvature of the gravitational N-body system is almost 
everywhere (\ie, almost always) positive even for moderately large $N$. 
In the very limit $N\gg 1$, $k_R(s)$ is always rigorously positive.
\item Well then, the geodesic instability too arise from the fluctuations of 
the {\sl positive} Ricci curvature, which in turn implies the fluctuations 
(\ie, the anisotropy) of the sectional curvatures of the manifold. 
In such a case, neither rigorous results nor convincing arguments exist 
compelling the instability time-scale, of the order of the dynamical one, 
to give a guess on the \SM relaxation time-scale.
\item The absence of any transition, analogous to the \SST~observed in the 
\FPU model, in the degree of chaoticity of the dynamics, leaves open the 
question of the existence of different regimes of rapidity in the approach 
to the equilibrium.
\item The amplitude of fluctuations and the level of anisotropy in the
$N$-body system are nevertheless amplified with respect to those occurring
in the \FPU chain. It is not hard to guess for the sources of the strong
anisotropy and of the big fluctuations in the configuration manifold.
The results obtained for few dimensional \DS's, \cite{CPMT_HH}, let us
to support the claim according to which a self gravitating $N$-body
system find itself in a regime of fully developed chaos, for the vast
majority of initial conditions, as long as $N\gg 1$.
\item The discussion reported above is the starting point for an 
investigation on the issue of the existence of {\sl one final} or 
{\sl various partial} \SM equilibria for $N$-body self gravitating systems. 
The analysis carried out on the relevance of the scaling law of the different 
terms appearing in the \GDI~ suggests a scenario with a full hierarchy of  
processes of approach to metastable equilibria, whose time scales are ordered 
in accordance to a criterion related to the scale (\ie, the fraction $n/N$ 
of the stars involved) of the process itself, and will appear soon.
\end{itemize}

\subsection*{Comparison between analytical estimates and numerical results.}
In the table \ref{tab_stime} we report the analytical estimates, slightly
improved with respect to those appearing in \cite{CPTD}, on the relevance 
and the $(\Ncal,\epsilon)$ dependence of the terms entering the \GDI, whose
full account will appear in \cite{PRE97b}, along
with some consistency checks obtained by numerical simulations. 
Minor corrections, which do not change the overall behaviours, have been
obtained taking into account a more careful statistical evaluation of 
the amplitude of those terms we named {\sl fluctuating}. 
As it is evident from the table, for the \fpub chain there are no doubt 
on the relative weights of various terms. When the virialization phase is 
completed, and this happen at once, when the number of oscillators is 
appreciably greater than few tens\footnote{
We remind that the virial equations in their general form: 
$\langle q^a \partial H/ \partial q^a\rangle = \langle p^a \partial H/ 
\partial p^a\rangle$, 
where the averages are taken over time or phase space, when applied to 
\mdf system, using the {\sl central limit theorem} and summing
over the degrees of freedom, let to state that 
the equality is true istantaneously with an approximation that becomes more
and more reliable as $N$ increases, being the statistical {\sl error} 
proportional to $N^{-b}$, where, depending on the interaction, $1\minord b
\minord 3$, see also note \ref{phasemix}.}, the average {\sl positive} 
values of both $Q(t)$ 
and $\hat{k}_R(t)$ are determined mostly by the laplacian term, while the 
fluctuations around these values, again after the first phases of approach 
to virial equilibrium, are due exclusively to the laplacian for $Q(t)$, 
whereas are influenced also by other terms for what concerns $\hat{k}_R(t)$. 
Nevertheless, although characterized by greater fluctuations, the average 
values of $\hat{k}_R(t)$ obviously coincide with $\langle Q(t)\rangle$, as 
shown in fig.\ref{fig_ricdiE}. 
Moreover, as shown in fig.\ref{fig_N-dilogt}, after the virial equilibrium has 
been attained, as long as we have $\Ncal\gg 1$ and the system is far from 
integrability, the probability of the occurence of a negative value of 
the Ricci curvature becomes actually zero.
Apart from this last remark, these results are substantially unaffected 
whether be the regime we are investigating, as all the terms scale with 
$\beta\epsilon$ in the same fashion, in such a way that their relative 
weights are unchanged. While, obviously, the amplitude of the fluctuations 
increases along with anharmonicity.\\
So, for the \fpub chain there aren't many cautions to be made, and the numerical 
simulations simply confirm the analytical estimations and support all the
argumentations based on them, \cite{CPPG_nc94,CPPG_Gen}. 
As a simple example we plot in figure \ref{fig_ricdiE}, from \cite{CPTD}, 
the $\epsilon$ dependences of $\langle Q\rangle$ and $\langle \hat{k}_R\rangle$, 
for three different values of $\beta$, which exploits neatly the foreseen 
scaling as $(\beta\epsilon)^{1/2}$.\\

As remarked above, the situation for the $N$-body problem is more involved, 
and the analytical estimates alone let us know only some of the answers, 
although perhaps the most physically relevant ones, namely, 
where come from the 
most important contribution to  $Q(t)$ {\it and} the scaling behaviours of 
the Ricci related quantities with the energy density $|\epsilon|$, which 
lead to the conclusion that all them depend on the energetic contents of 
the system exactly as they have to do; for example, that the average value, 
measured in {\sl natural units}, 
$\langle Q\rangle\, t_D^2$ {\sl do not depends on} $\epsilon$. 
Stated otherwhise, this indicates that 
the scale invariance of gravity manifests itself within geometrical 
transcription in such a way that all the (possibly suitably rescaled to meet 
dimensional needs) averages of \GDI~have the same values if measured in 
the unit of dynamical time scale $t_D$.\\
Moreover, having singled out the dominant contributions to $Q(t)$, we are 
able to definitely support our claims that even in this case is the 
{\sl parametric resonance} to drive to the onset of Chaos, and that the
system dynamics should undergo a transition along the approach to the
virial, if the initial conditions are chosen out of equilibrium.\\
What it is left to determine is the scaling law behaviour of the \GDI~with 
$N$, which is practically absent in {\sl extensive} potentials, but 
nevertheless likely to exist in the  {\sl superextensive}
N-body problem, as indeed shown in table \ref{tab_stime}.
The correct behaviour has been established supplementing the analytical 
investigations with numerical simulations whose details will be 
presented elsewhere, but whose main outcomes are:
\begin{itemize}
\item The {\sl frequency} $Q(t)$ is in almost always positive for the 
N-body problem, and the chances to find a negative value decrease 
quickly when $N$ increases. 
To show this, we report in figures \ref{fig_Qdit10}, \ref{fig_Qdit100} 
and \ref{fig_Qdit200} the time evolution of all the contributions coming 
from {\sl static} and {\sl fluctuating} terms.
\item Despite the positive values of both $Q(t)$ and $\hat{k}_R(t)$ 
(whose average is always positive and for which also the occurrence of 
local negative values becomes negligibly small when $N$ increases) imply 
that also the Ricci curvature per degree of freedom, $k_R(s)$, is positive; 
we found that the dynamics of self gravitating $N$-body systems is always 
strongly chaotic, \cite{PRE97b}, as far as chaoticity can be defined 
for non-compact phase spaces. This is explained again by the phenomenon of 
parametric resonance 
which is always in the regime of fully developed
instability, as the fluctuations in the \GDI~ are for this \DS~constantly
above the threshold required for the mechanism to onset.
\item All quantities entering the \GDI~have the same 
specific energy $\epsilon$ dependence 
as $t_D^{-2}$, \ie, both the laplacian and gradient related 
quantities, on one side, and the first (squared) and second time derivatives 
of the total 
potential energy, on the other, scale exactly
as $|\epsilon|^3$, in such a way that the {\sl frequencies} they define are in 
constant ratio with the {\sl natural} orbital one.
\item The $N$ dependence of these quantities, even when removed that coming 
from the dynamical time (we recall that
we have $t_D\propto N|\epsilon|^{-3/2}$), manifests 
a very peculiar behaviour, with all quantities changing with $N$,
 but in such a way that the hierarchy of their relative ratios either
remains unchanged or becomes even more pronounced; figure \ref{fig_GDIdiN}
shows the behaviour of the {\sl contributors} to the Ricci related
frequencies when $N$ is varied.
\item With the support of specifically devoted numerical simulations, we 
explained this peculiar $N$-dependence, which turns out to be very enlightening 
also on the consequences of the {\sl gravitational clustering}, whose basic 
more elementary example is the formation of a binary system. 
The implications of this point on the evolutionary processes of stellar systems
are currently under study.
\item The numerical simulations performed confirmed that the 
analytical estimates correctly describe the behaviour with $N$ of the ratios 
between the terms containing the 
time-derivatives of the kinetic energy of the system.
\item Also for the {\sl static} terms the {\it a priori} estimates correctly 
indicate their relative weights and even their ratios to the {\sl fluctuating} 
ones.
\item We investigated in detail also the behaviour of the {\sl static} terms, 
and interpreted their behaviour with $N$ on the basis of an analytical 
conjecture, confirmed by numerical computations, whose details will
be discussed in \cite{PRE97b}, where we will show also how
 all these outcomes do not depend on the detailed way we
adopted to cope with the mathematical singularity related to the laplacian 
term in order to restore its physically relevant role.
\end{itemize}
Figure \ref{fig_GDIdiN} clearly shows the precise scaling law of all 
the quantities entering the \GDI~ on $N$, along with the trivial
dependence on $|\epsilon|$, or, is is the same, on $t_D$.
There are other minor comments which can be made to the results of numerical 
simulations, among which we mention
the very small
oscillations in the panels {\tt a)} and {\tt c)} of figure \ref{fig_N-dit},
 along with their long overall {\sl quasi-periodicity} 
(the time is in {\it millions} of harmonic units!). 
Moreover, panels {\tt b)} and {\tt d)} of the same figure point out how
the relationship between Chaos and frequency of a negative
sign of the Ricci curvature should be at least, {\sl bimodal}.

\section*{Conclusion.}
This paper belongs to a line of research addressed to the issue of a 
{\sl geometric description of Chaos} in \DS's. Here we focused our attention 
mainly of \mdf hamiltonian systems, in order to investigate the sources of 
instability and to single out possible {\it a priori} criteria to detect the 
onset of chaotic behaviour, in analogy with what obtained in the case of few 
dimensional \DS's, \cite{CPMT_HH}.\\
We discussed how the criteria borrowed from abstract Ergodic theory could be 
used, if ever a realistic physical model would fulfil their rather restrictive 
hypotheses. To this goal we exhaustively investigated, starting at a very 
basic level, the relationships existing between the various {\sl curvatures} 
that can be defined over a manifold {\sf M} where the geodesic flow 
translating the dynamics in a geometrical picture takes place.\ We derived some 
of their logical consequences, and then indicated some cautious remarks
in order to get reliable indications from ambitious applications of the \GDA, 
\cite{GS,Kandrup}.\\
In particular, we here reported the formal explanation (see,\eg, \cite{CPTD}), 
within a more general context, of the analytical and numerical results we 
obtained, \cite{CPPG_nc94,MTCP_BIX,CPMT_HH}, about the irrelevance of scalar 
curvature and related quantities to detect the presence of instability,
 obviously as long as $\Ncal\geq 3$.\\
It is only recently, \eg, \cite{Marco93,CPTD}, that a careful framework has 
started to be set up. Thanks to this more systematic approach, the method has 
revealed, and we claim only in part, its fruitfulness, for both \mdf and 
few dimensional \DS's.\\
In the present work we exploited its ability to single out the transition  
between two different regimes of Chaos in the \fpub model. Further, we also 
reinterpreted some sources of misunderstandings 
for what concerns the study of instability in the gravitational N-body problem.
\ We shown its reliability performing analytical calculations, aided and 
supported by numerical computations, able to detect the scaling law behaviour 
of relevant quantities entering the equations governing the second order 
variational equations and in such a way identifying once more the peculiar 
behaviour on Newtonian Gravity.
We proved nevertheless that the mechanisms responsible of the onset of 
Chaos in {\sl standard}
\mdf Hamitonian systems of interest in a classical Statistical Mechanics 
context, are also at work to determine the {\sl exponential instability} 
(if not {\it true Chaos}), in this {\sl not well behaved} \DS.\ In this way, 
the somewhere persistent claims, \cite{GS,Kandrup}, about a na\"\i ve 
relationship between instability and {\sl relaxation} time scales should be 
revisited, and that their identification should be considered definitely false. 
Rather, it has to be reminded that for some \DS's there are some analogies
in their behaviours, though limited to the existence of a
{\sl common transition} between two different regimes, \cite{Marco91}, 
but that, for the N-body system no relationship has any more than 
heuristic support, \cite{CPPG_nc94,CPPG_Gen}.\\
The \GDA~has made possible, the {\it analytical} 
{\sl evaluation of the maximal} \LCN for the \FPU chain, through simple 
procedures developed in the framework of stochastic differential equations, 
\cite{Vankampen},using solely the time average of the Ricci curvature and 
of its dispersion 
along with an estimate of its autocorrelation time 
(see \cite{CPTD,PRE97b} and \cite{MarcoLapo}).
The analytically calculated maximal \lcn's are in a definitely satisfactory 
quantitative agreement with those numerically computed using the standard 
tangent dynamics equations, except at a very low (specific) energy, \ie, 
when Chaos is not yet fully developed, as in this case {\sl the fluctuations} 
of \GDI~ show a slow convergence to their asymptotic values (whereas the 
average values of the curvatures are reached very quickly also in the harmonic 
regime). Incidentally, this is a signature of the many faces of the approach 
to {\sl the equilibrium}, which is a concept strongly dependent on the 
observable you are looking to {\sl relax} to its
{\sl ergodic value}. Recalling a previous discussion, we furthermore observe 
that similar results are obtained averaging over phase space, instead of on time, 
and it should be intriguing to ask how the \GDA~allows also to gain some 
hints on the relaxation 
time scales for \mdf systems whose final equilibrium is guaranteed to exist.\\ 
The framework outlined here and previously developed in \cite{CPTD,MTTD} 
serves as a starting point to try to answer also these basic questions. 
In the near future we will exploit 
 the results obtained both in the case of few dimensional \DS's
 and for \mdf lagrangian systems, 
to shed light, from a novel perspective, to the long standing issue of a 
dynamical justification of the statistical approach in (classical) 
mechanics.

\subsubsection*{Acknowledgements:} Both authors are indebted to Marco Pettini 
for the unselfish encouragements given in all circumstances; 
P.C. thanks Francesco Paolo Ricci and Michele Nardone, that allowed him 
to come back occasionally from molecular dynamics potentials to 
gravitational interaction.\\
The first part of this work started when both the authors
were at the {\sf G9} group of Physics Dept. at the University 
of Rome {\it "La Sapienza"}. 

\begin{center}
{\Large {\bf Figure Captions}}
\end{center}
\begin{figure}[h]
\caption{Long time behaviour of the measured {\sl cumulative frequency}
 of occurrence
of negative values of the Ricci curvature $F_{-}(t)$ in the \fpub~chain,
for a set of initial conditions {\sl far enough} from integrability. 
All the solid lines refer to $N=450$ coupled oscillators with 
 $\beta\epsilon$ ranging from $5\cdot 10^{-3}$ to $10^3$.
The dashed lines represent the behaviour of a smaller system ($N=150$)
not too far from integrability, $\beta\epsilon=10^{-3},\ 10^{-2}$, from 
bottom to top, showing the greater noise affecting simulations
with too small $N$.
\label{fig_N-dilogt}}\vskip.25cm

\caption{The same of the figure \ref{fig_N-dilogt}, but in the opposite
situation: very small $N$ and/or conditions very near to integrability. 
The values of $(N,\beta\epsilon)$ are: 
{\tt a)} $(50,5\cdot 10^{-5})$; 
{\tt b)} $(50,0.03)$ (upper curve) and $(50,0.01)$ (lower curve);
{\tt c)} $(50,5\cdot 10^{-4})$;
{\tt d)}, from top to bottom, $(150,5\cdot 10^{-5})$, $(150,10^{-4})$, 
$(450,5\cdot 10^{-5})$ and $(150,2\cdot 10^{-4})$. 
The horizontal scale now is always linear and time is there expressed in million
of units.
\label{fig_N-dit}}\vskip.25cm

\caption{Energy dependence of the largest LCN of the \fpub~chain,
computed in analytical way. The dashed lines trace the low and high 
energy slopes: $(\beta\epsilon)^2$
and $(\beta\epsilon)^{1/4}$, respectively. In the small panel 
we plotted the {\sl asymptotic} values of the cumulative 
frequency $F_{-}(t)$ for the same conditions used to compute the LCN's. 
\label{fig_N-vsLyap}}\vskip.25cm

\caption{Time averages of the {\sl rescaled} Ricci curvature 
$\hat{k}_R(t)$ and of the {\sl Hill frequency} $Q(t)$ for the same set
of initial conditions use to compute the largest LCN. The continuous
line is the plot of the analytic estimate reported in 
table \ref{tab_stime}. The {\sf SST} is neatly visible.
\label{fig_ricdiE}}\vskip.25cm

\caption{Time behaviour of the quantitities entering the {\sf GDI}'s, 
$Q(t)$ or $\hat{k}_R(t)$, for the gravitational N-body system with $N=10$ 
(!) and $|\epsilon|=1$. The small panels show, from left to right,
the contributions to $Q(t)$ (shown alone in the larger window)
 from the {\sl fluctuating} and {\sl static} terms,
respectively. In the small panels here and in the following figures, 
the laplacian term is indicated with a solid line, 
the squared gradient as dot-dashed, the (positive values) of the 
second derivative of the potential energy as dashed, 
and the dotted curve display its squared first time derivative.
In this and the following figures all the terms are 
plotted with the weights they enter in the expression of $Q(t)$.
\label{fig_Qdit10}}\vskip.25cm

\caption{The same as figure \ref{fig_Qdit10}, but for $N=100$. 
 Note the logaritmic vertical scale of the inserted plot. The quantities
are represented as in the previous figure.
\label{fig_Qdit100}}\vskip.25cm

\caption{The same as the previous two figures but for $N=200$ 
and with all the quantities in a plot. The small solid triangles trace 
the behaviour of $Q(t)$ in order to try to disentangle it from that of 
$\Delta\Ucal/\Ncal$, with which practically coincides. 
The other lines confirm the relative hierarchy of other terms. 
\label{fig_Qdit200}}\vskip.25cm

\caption{In the large area we plot the $N$ dependence of the time 
averages of the various terms. These different scaling with $N$ 
are explictly shown in the inserted plot, which also points out the 
{\sl sharp} $|\epsilon|^3$-dependence of all these quantities.
\label{fig_GDIdiN}}
\end{figure}
\hoffset-3.truecm
\begin{center}
{\Large {\bf Tables}}
\renewcommand{\arraystretch}{1.2}
\begin{table}[h]
\caption{Analytical estimates of the $(N, \epsilon)$-dependence of
quantities entering the {\sf GDI}'s and of the {\sl dynamical} 
 time, for the {\sf FPU}-$\beta$ chain and
the Gravitational N-body system, along with the indication about
the scaling law behaviours {\sl observed} for the instability exponents 
of both dynamical systems, evaluated either numerically or using the 
semi-analytical method discussed in the text. 
In this table $N$ represents the number of oscillators or the
number of {\sl masses}. We remark that the amplitudes here
indicated should be taken as {\sl order of magnitude estimates}, up to
numerical factors of order unity. Moreover, for what concerns the
{\sl fluctuating terms}, these evaluations represent almost the 
largest absolute value,
as they average can be much smaller (even vanishing!). The quantities
indicated with the symbols $C_{F_n}$ depend on the clustering level of the
N-body system and then slowly evolve, $\eta$ is a tracer of the way we
adopted to cope with the newtonian singularity and the exponents $a_n$ reflect
the efficiency of the {\sl virial damping} on the amplitudes of 
evolving terms. Their explicit expression, as well that of the 
analytical formula, $\Lambda$, to compute the LCN, is given in 
\cite{CPTD,PRE97b}, see also \cite{MarcoLapo}.
\label{tab_stime}}
\vskip.8cm
{
\begin{tabular}
{||c||l|l|c||l|c||}\hline
\              & \multicolumn{3}{c||} {\large{\bf FPU}}               & 
                                               \multicolumn{2}{c||} 
{\large{\bf N-Body}}\\ \hline
{\sf Quantity} & {\sf Amplitude} & {\sf Limiting Behaviour}
$\begin{array}{l}
\beta\epsilon \ll 1\\
\beta\epsilon \gg 1
\end{array}$
& {\sf VDF} & {\sf Amplitude}   & {\sf VDF}\\ 
\hline
$ {\displaystyle\frac{\Delta{\cal U}}{N} } $ & 
$ 2 + 6 \left(\sqrt{1+\beta\epsilon}-1\right) $ &
$
\begin{array}{l}
2+ 3\beta\epsilon +{\cal O}(\beta\epsilon)^2\\
6 \sqrt{\beta\epsilon}\ + {\cal O} (\beta\epsilon)^{-1/2}
\end{array} $
 & ${\cal O} (1)$ & 
${\displaystyle
\eta^2 {{|\epsilon|^3}\over{ N^2}} C_{F_5}
}$
 & $ {\cal O} (1) $ \\ \hline

${\displaystyle
{{(\grad{\cal U})^2}\over {N\Wcal}} 
}$ & 
$(1 + \beta\epsilon) {\displaystyle
 {{\sqrt{1+\beta\epsilon}-1}\over {N\beta\epsilon}} }$
 &
$
\begin{array}{l}
(1+ \beta\epsilon) \ /\ 2N\\
\sqrt{\beta\epsilon} /N
\end{array} $
 & ${\cal O} (1)$ & 
${\displaystyle
{{|\epsilon|^3}\over{ N^4}} C_{F_4}
}$
 &$ {\cal O} (N^{a_3})$\\ \hline

${\displaystyle
{{|\ddot{\cal U}|}\over {N\Wcal}} 
}$ & 
$
(1 + \beta\epsilon) {\displaystyle
 {{\sqrt{1+\beta\epsilon}-1}\over {N^{3/2}\beta\epsilon}} }
$
 &
$
\begin{array}{l}
\null \\
\sim N^{-1/2}{\displaystyle
{{(\grad{\cal U})^2}\over {N\Wcal}} 
}\\
\null \\
\end{array} $
 & ${\cal O} (N^{a_1})$ & 
${\displaystyle
 {{|\epsilon|^3}\over{ N^{4}}} C_{F_4}
}$
 &$ {\cal O} (N^{a_2})$\\ \hline

${\displaystyle
{{1}\over{N}}\left({{\dot{\cal U}}\over{\Wcal}}\right)^2 
}$ & 
$
(1 + \beta\epsilon) {\displaystyle
 {{\sqrt{1+\beta\epsilon}-1}\over {N^2\beta\epsilon}} }
$
 &
$
\begin{array}{l}
\null \\
\sim N^{-1}{\displaystyle
{{(\grad{\cal U})^2}\over {N\Wcal}} 
}\\
\null \\
\end{array} $
 & ${\cal O} (N^{2a_1})$ & 
${\displaystyle
 {{|\epsilon|^3}\over{ N^{5}}} C_{F_4}
}$
 &$ {\cal O} (N^{2a_2})$\\ \hline

$t_D$ & 
$
{\displaystyle
\left({{\sqrt{1 + 4\beta\epsilon} -1}\over {2\beta\epsilon}}\right)^{1/2}
}
$
 &
$
\begin{array}{l}
1- \beta\epsilon/2 +{\cal O}(\beta\epsilon)^2\\
(\beta\epsilon)^{-1/4}\ + {\cal O} (\beta\epsilon)^{-3/4}
\end{array} $
 & --
 & 
${\displaystyle 
{{N}\over {|\epsilon|^{3/2}}}
}$
 &-- \\ \hline

$LCN$ & 
$
\Lambda\, (\bar{\chi},\sigma_\chi,\tau_\chi)
$
 &
$
\begin{array}{l}
(\beta\epsilon)^2\\
(\beta\epsilon)^{1/4}
\end{array} $
 & --
 & 
${\displaystyle 
{{|\epsilon|^{3/2}}\over {N}}
}$
 &-- \\ \hline
\end{tabular}
}     
\end{table}
\end{center}
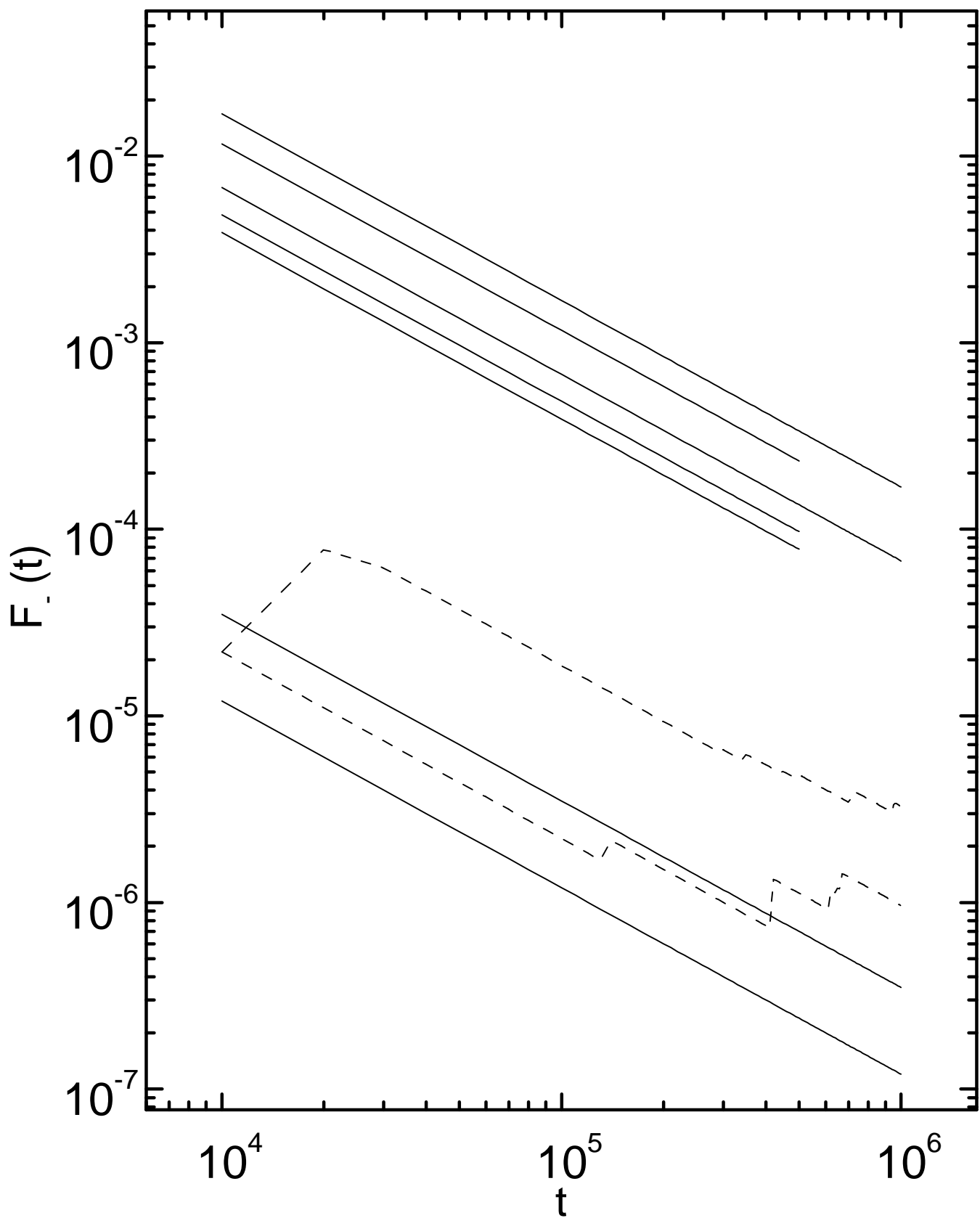
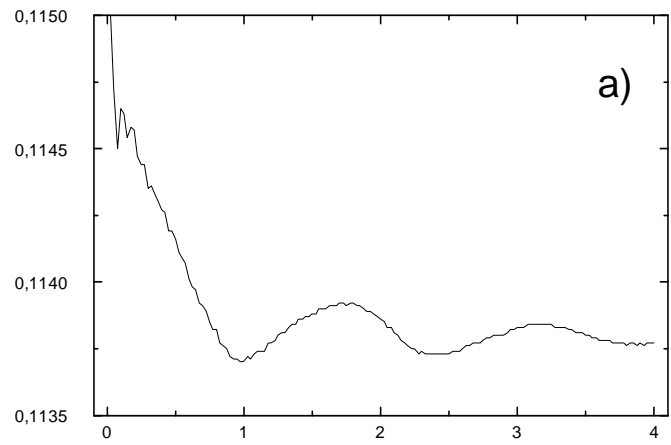
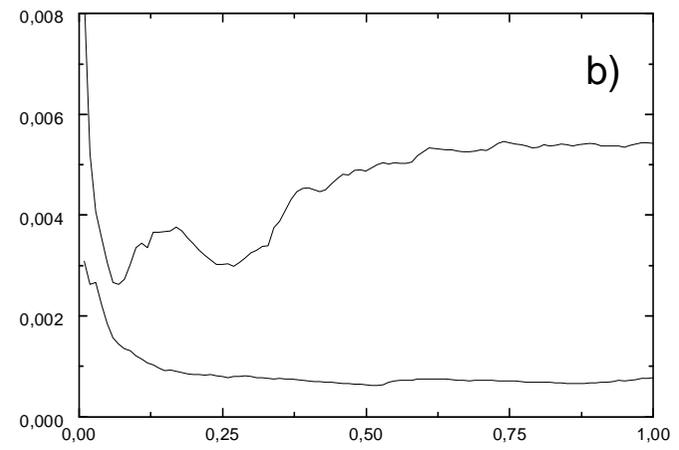
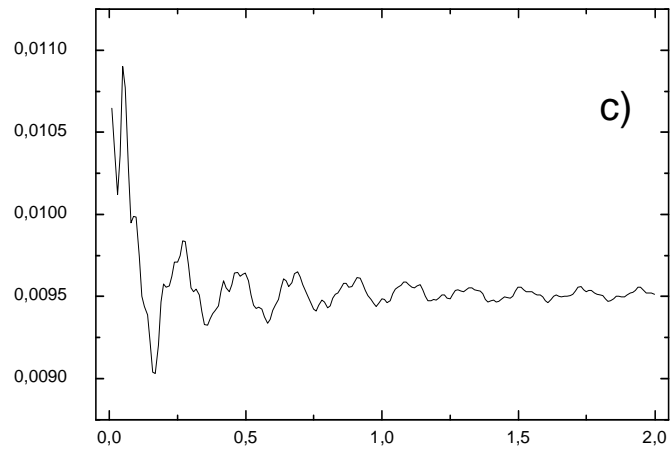
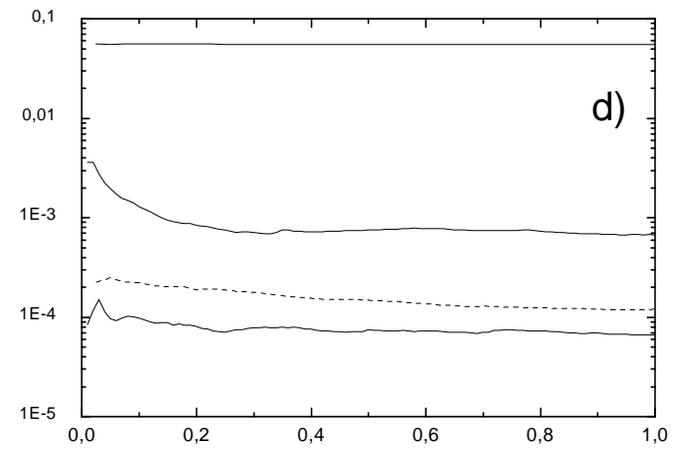
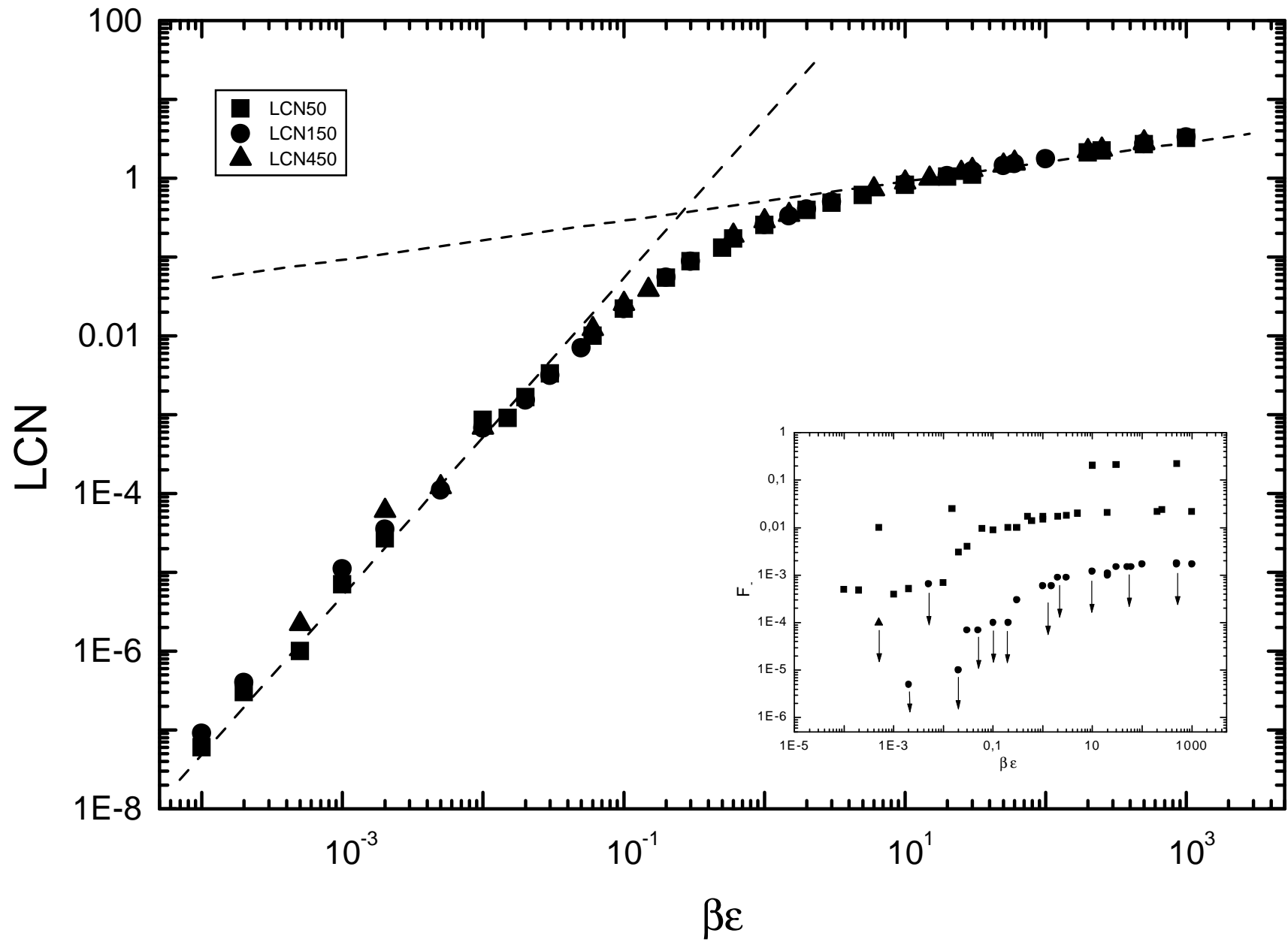
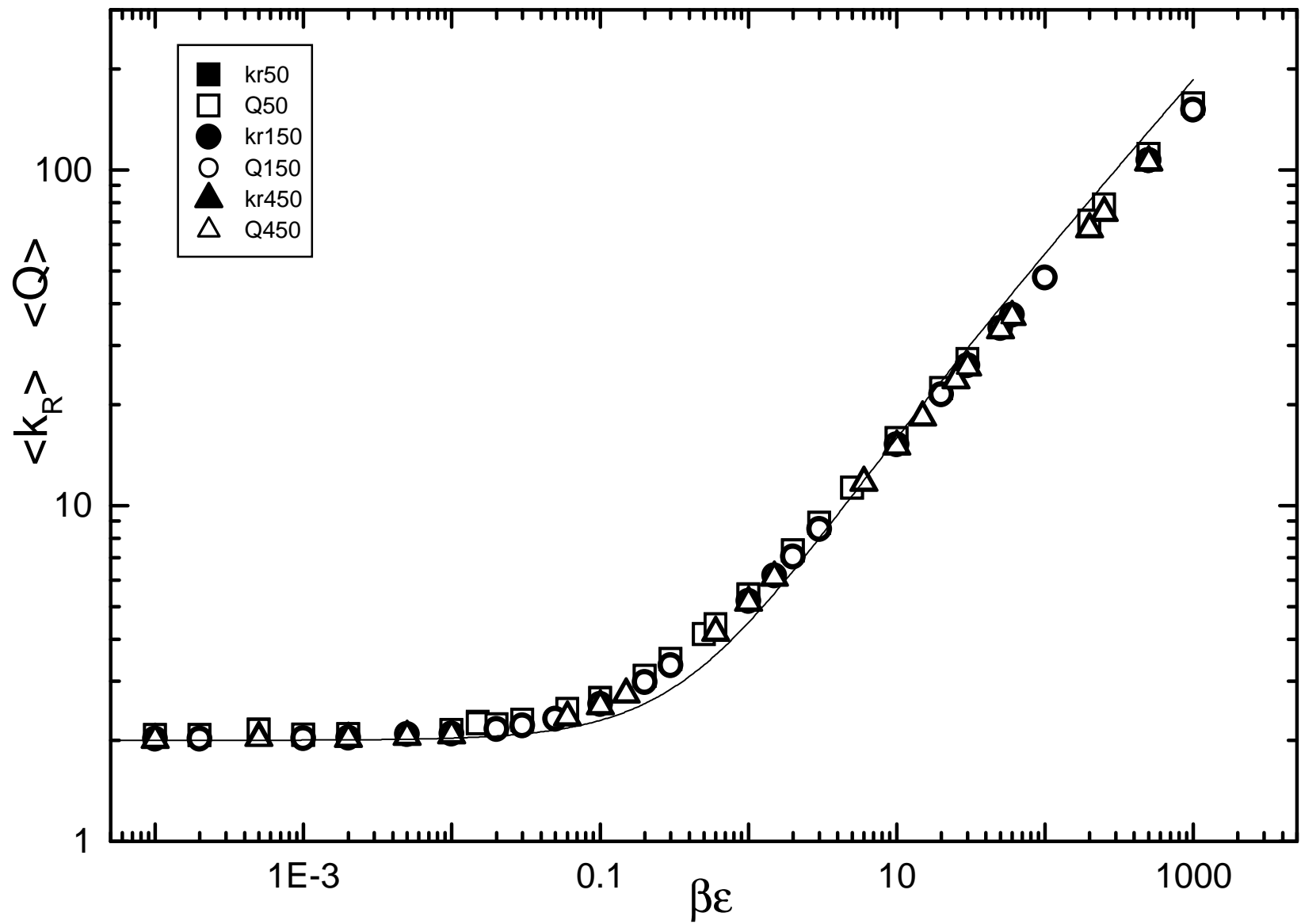
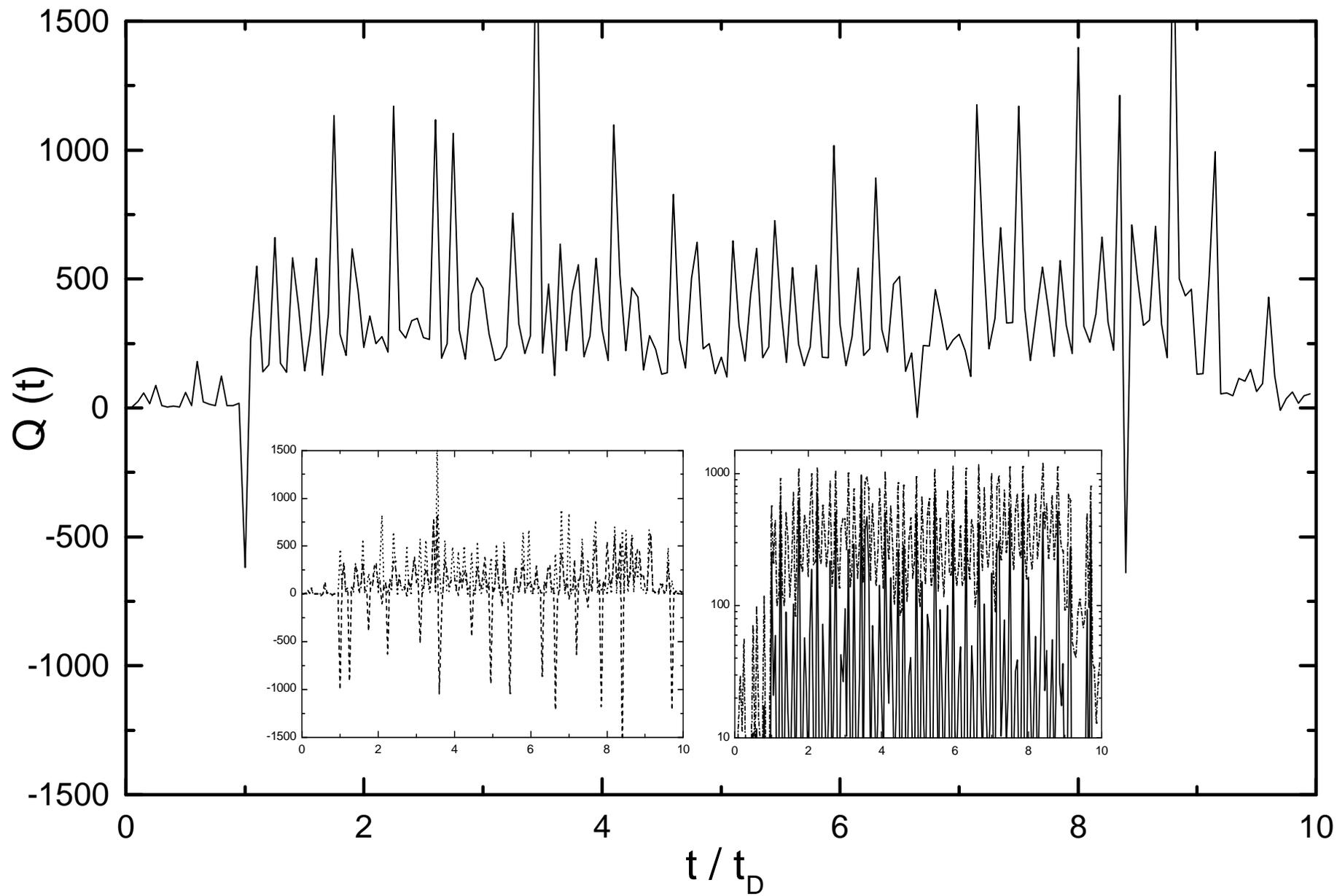
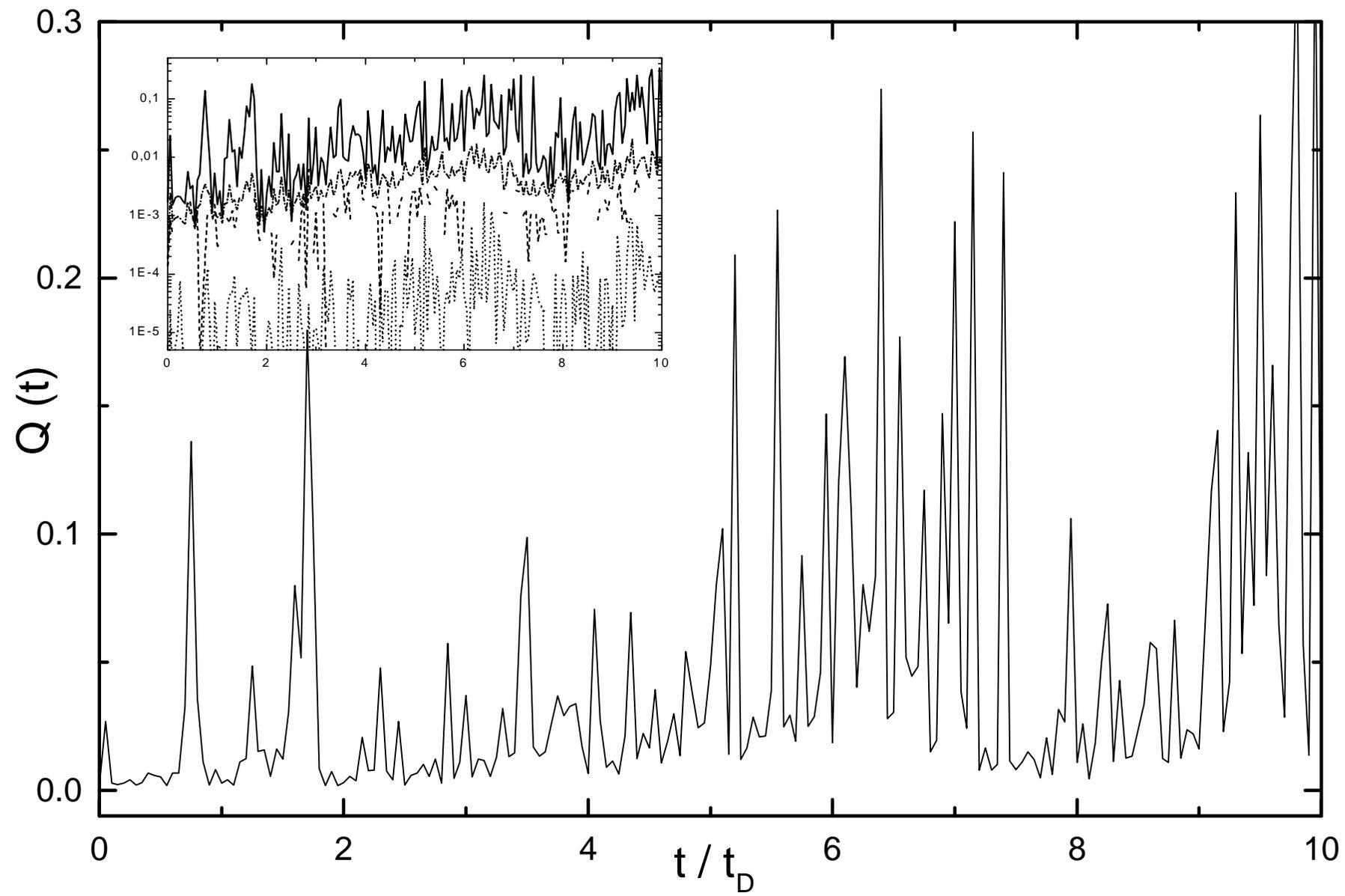
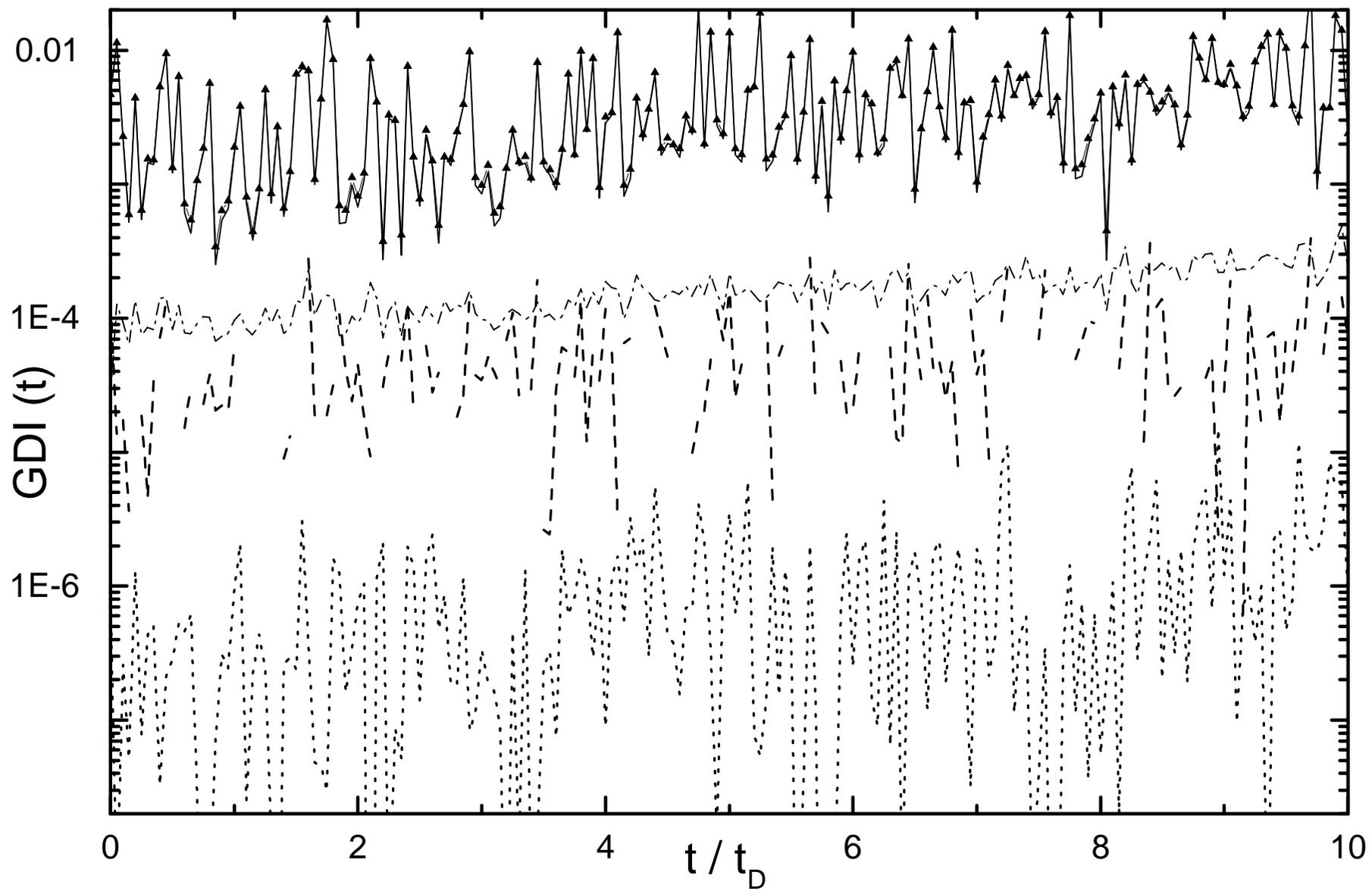
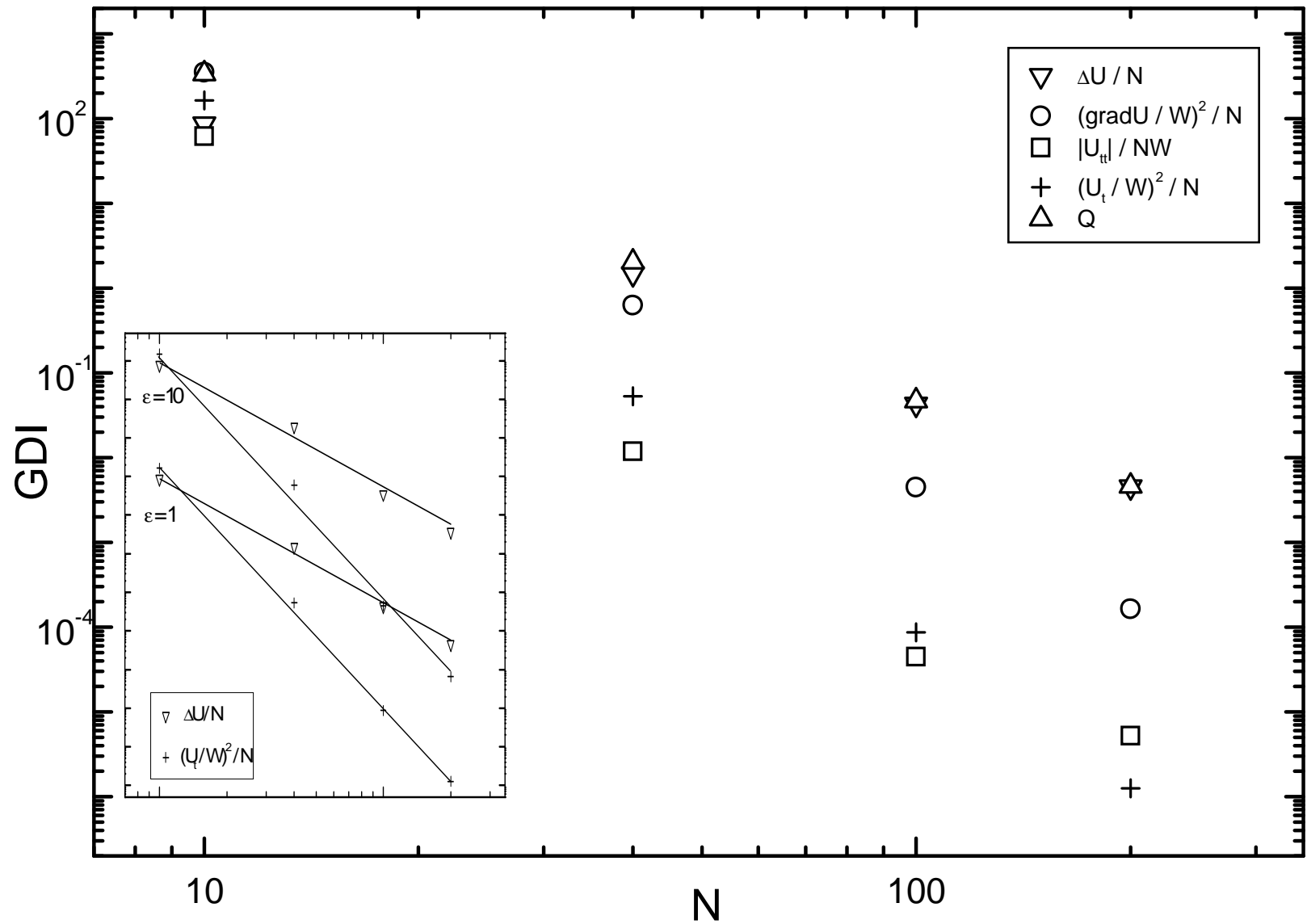
\end{document}